\documentclass[10pt, a4paper]{article}
\usepackage{moreverb}
\usepackage{authblk} 
\usepackage[colorlinks,bookmarksopen,bookmarksnumbered,citecolor=red,urlcolor=red]{hyperref}

\usepackage[svgnames, x11names]{xcolor}
\definecolor{RED}{rgb}{0.8, 0.0, 0.0}
\definecolor{MAGENTA}{rgb}{0.55, 0.0, 0.55}
\definecolor{ORANGE}{rgb}{0.8, 0.33, 0.0}
\definecolor{BLUE}{rgb}{0.0, 0.0, 0.8}

\newcommand{\ysnoted}[1]{}

\usepackage[normalem]{ulem} 

\usepackage{listings}

\lstdefinelanguage{XML}
{
basicstyle=\ttfamily\footnotesize,
  morestring=[b]",
  moredelim=[s][\bfseries\color{Maroon}]{<}{\ },
  moredelim=[s][\bfseries\color{Maroon}]{</}{>},
  moredelim=[l][\bfseries\color{Maroon}]{/>},
  moredelim=[l][\bfseries\color{Maroon}]{>},
  morecomment=[s]{<?}{?>},
  morecomment=[s]{<!--}{-->},
  commentstyle=\color{gray},
  stringstyle=\color{blue},
  identifierstyle=\color{red}
}
\usepackage[pdftex]{graphicx}
\graphicspath{{./../}}
\DeclareGraphicsExtensions{.pdf}

\usepackage{amssymb}
\usepackage{mathtools}

\usepackage[caption=false,font=scriptsize]{subfig}
\graphicspath{{./figures/}}

\usepackage{algorithmicx}
\usepackage{algpseudocode}
\usepackage[ruled]{algorithm}
\definecolor{light-gray}{gray}{0.75}
\algrenewcommand{\algorithmiccomment}[1]{\hskip3em{{\footnotesize \textcolor{light-gray}{$\blacktriangleright$}}} #1}

\usepackage{multirow}
\usepackage{rotating}

\newcommand{\ssec}[1]{{\emph{$\circ$~#1:~}}}

\usepackage{hyperref} 

\usepackage{xspace}

\usepackage{enumitem}

\begin{document}

\title{Characterizing Application Scheduling on\\ Edge, Fog and Cloud Computing Resources\thanks{Pre-print version of article to appear: Varshney P, Simmhan Y. Characterizing application scheduling on edge, fog, and cloud computing resources. \emph{Softw: Pract Exper.} 2019; 1--37. \url{https://doi.org/10.1002/spe.2699}}}

\author{Prateeksha Varshney}
\author{Yogesh Simmhan}
\affil{Indian Institute of Science, Bangalore, India}
\affil{\em Email: prateekshav@iisc.ac.in, simmhan@iisc.ac.in}

\maketitle

\begin{abstract}
Cloud computing has grown to become a popular distributed computing service offered by commercial providers. More recently, Edge and Fog computing resources have emerged on the wide-area network as part of Internet of Things (IoT) deployments. These three resource abstraction layers are complementary, and offer distinctive benefits. Scheduling applications on clouds has been an active area of research, with workflow and dataflow models offering a flexible abstraction to specify applications for execution. However, the application programming and scheduling models for edge and fog are still maturing, and can benefit from learnings on cloud resources. At the same time, there is also value in using these resources cohesively for application execution. 
In this article, we offer a taxonomy of concepts essential for specifying and solving the problem of scheduling applications on edge, for and cloud
computing resources. We first characterize the resource capabilities and limitations of these infrastructure,
and offer a taxonomy of application models, Quality of Service (QoS) constraints and goals, and scheduling techniques, based on a literature review. We also tabulate key research prototypes and papers using this taxonomy. This survey benefits developers and researchers on these distributed resources in designing and categorizing their applications, selecting the relevant computing abstraction(s), and developing or selecting the appropriate scheduling algorithm. It also highlights gaps in literature where open problems remain.
\end{abstract}

\section{Introduction}

In computing, \emph{scheduling} refers to the process of allocating computing resources to an application and mapping constituent components of that application onto those resources, in order to meet certain Quality of Service (QoS) and resource conservation goals~\cite{scheduling-book}. The \emph{application} itself may be represented in an abstract or concrete form using different programming primitives such as processes, threads, tasks, jobs, workflows, petri nets, and so on~\cite{process-thread-sched,job-sched,petri-sched,gannon-wfbook}. Similarly, the \emph{computing resource} may be diverse, ranging from local cores and processors on a host, to distributed resource like nodes in a cluster, virtual machines (VM) in a cloud, edge or mobile devices in an Internet of Things (IoT) deployment, or desktops in a volunteer computing network~\cite{hpc-sched,grid-survey2,zhan:csur:2015,8123913,BOINC}. \emph{QoS} for the application, such as their latency, and \emph{conservation goals}, such as minimizing the quanta of resource or their energy footprint, can likewise be used to determine the schedule. Consequently, examining application scheduling requires us to understand the behavior of the computing resources, application models, and QoS goals in an integrated manner.

There is a growing availability of heterogeneous distributed computing resources. \emph{Cloud computing} is a capability offered by commercial service providers using a rental model. Here, virtualized compute and storage resources at large data center with thousands of servers are available on-demand~\cite{berkeley:2009}. In addition, there has been the emergence of devices at the \emph{Edge} of the network as part of the broader roll-out of \emph{Internet of Things (IoT)}. These can be sensors and devices that are part of Smart City infrastructure, lifestyle gadgets like wearables, and smart appliances or smart phones. Besides sensing and generating observation streams, these devices that number in the tens of thousands have spare compute, storage and memory capacities. These can be leveraged to execute IoT applications at the edge of the network~\cite{ghosh2018adaptive}. Further, there has been heightened interest in \emph{Fog resources}, that are between the edge and cloud in the network hierarchy, with compute, storage and memory capacities that fall between these layers as well~\cite{simmhan2018fog}. These edge and fog resources provide the opportunity for low latency processing of the generated data, closer to its source, and on the wide-area network~\cite{stojmenovic:2014}.
As a result, there is critical need to understand how this diverse ecosystem of edge, fog and cloud resources can be effectively used by large-scale distributed applications.

\emph{Scheduling applications on edge, fog and cloud} deals with the placement of the application's logic components onto these resources for execution, deciding their interactions within this compute and network hierarchy, and managing various forms of dynamism, to meet their QoS requirements. There can be resource mobility at the edge and fog layers. The applications may also impose requirements on logical mobility of processes and data. Further, there may be changes in the data generation rates, network behavior or energy levels of batteries that require reactive strategies~\cite{shi2012serendipity}. The application needs to be scheduled and coordinated in order to meet various QoS goals such as latency, energy and monetary constraints. As a result, scheduling within this complex ecosystem involves a multitude of online coordination and optimization decisions, and the flow of control signals and data that can impact the application performance and resource usage.

There has been substantial work on examining scheduling approaches on clouds and clusters~\cite{zhan:csur:2015,kessaci:ipdpsw:2014}. However, given the nascency of edge and fog resources, \emph{there is a lack of a systematic review of distributing and scheduling applications on these resource classes individually, and together with clouds~\cite{hong:mcc:2013}.} 
\emph{Existing literature} has proposed the conceptual foundations of edge and fog computing~\cite{bonomi2012Fog,dasterdi:corr:2016,chiang-2016,simmhan2018fog}.
Others, including us, have discussed the benefits and challenges involved in the coordination among \emph{edge, fog and cloud layers} in a hierarchical model~\cite{7721750,varshney2017icfec}, but fail to examine in detail their impact of the applications and their schedule.
Several scheduling approaches also do not distinguish between the edge and the fog layers, and subsume the former into the latter (or vice versa)~\cite{skarlatprovisioning,skarlat2017towards,deng2016optimal}. There is also divergence in the assumptions made on the reliability and costing for the edge and fog layers. Hence, there is the need to understand the possible architectural patterns and scheduling mechanisms that have been proposed to cohesively schedule on these resource abstraction layers to inform researchers on open problems, and developers on available approaches and their relative merits.

\emph{A literature survey} can offer a framework to examine and understand such fast-paced emerging research, in the context of existing works. Preliminary surveys on \emph{Mobile Edge Computing (MEC)}, review task offloading strategies adopted by \emph{mobile} edge devices that coexist with cloud resources, motivated by the growth in smart phones~\cite{8030322,7879258}. However, these tend to focus on scheduling individual smart phone apps on a single edge device and the cloud. We generalize beyond mobile edges to all types of edge resource, include fog computing as a first-class entity, and consider diverse application and scheduling models on them. Some of these also consider the evolution of mobile edge computing with the advent of 5G communication technologies, with capabilities like network slicing and network function virtualization~\cite{taleb2018survey}. We approach this survey from the application and Infrastructure as a Service (IaaS) perspective, rather than examine the internals of the hardware or communications architectures.
Others have also summarized the relative characteristics of the three resource layers, similar to the System Design branch of our taxonomy~\cite{8123913}. However, they do not examine the impact of this on application design and scheduling models.

We distinguish our work from numerous cloud computing surveys, that review \emph{the breadth of the cloud ecosystem}~\cite{zhan:csur:2015}, \emph{scheduling of VMs} onto hosts~\cite{kessaci:ipdpsw:2014,mann:csur:2015}, and use of multi-cloud environments~\cite{grozev:spe:2014}. These are at different levels of abstractions, even within cloud computing. 
We also contrast with specialized reviews on specific scheduling techniques or resource feature, such as meta-heuristics~\cite{wu:2015,tsai:jsyst:2014}, elasticity, and fault-tolerance~\cite{coutinho:2015,huang:jsw:2013}. Rather, we take a holistic view of these and other characteristics such as pricing, variable performance, and application models, and consider them in the presence of edge and fog resources as well. 
the application and system models as well. 
These related works are reviewed in greater detail in Section~\ref{sec:related}.

We address these gaps and present a \emph{survey on scheduling of applications on to Edge, Fog and Cloud computing resources}, both independently and collectively, based on a review of contemporary scheduling literature. 
Specifically, we present a \textbf{taxonomy of concepts and approaches} for scheduling applications on edge, fog and cloud resources~(\S~\ref{sec:taxonomy}), based on a detailed review of research literature over the past decade. This classification covers \emph{properties of edge, fog and cloud resources} relevant to scheduling, \emph{characteristics of the application} that impacts the schedule, and the \emph{diverse QoS and constraints} that determine the performance of the schedule, and \emph{categories of scheduling algorithms} that exist. We then \textbf{tabulate key scheduling literature} using this taxonomy to offer a birds-eye view of the landscape of application scheduling on these resources ~(\S~\ref{sec:table}). We place our survey in the context of other \textbf{related surveys} that exist, and argue its novelty and impact~(\S~\ref{sec:related}). Lastly, we discuss \textbf{emerging trends} in this decade-old research area, and highlight open problems that the research community is actively exploring at present~(\S~\ref{sec:discussion}).

At the same time, our \emph{goal is not} to examine specific implementations of edge or fog computing technologies, cloud service offerings beyond IaaS (and even that with an emphasis on computing resources), nor to offer case-studies of applications. We compare and contrast the conceptual features across these resources, and abstract the higher-order application models to help examine scheduling techniques. We also do not consider security and privacy aspects such as authentication, encryption and cyber-attacks on edge, fog and cloud. Networking and communications technologies, and data center management are out of scope as well. Existing literature, some of which we review in the related work, address these adequately.

This article draws on \emph{our two prior works} that characterize the resource behavior of edge, fog and cloud~\cite{varshney2017icfec}, and offer an overview of fog computing~\cite{simmhan2018fog}. These content are selectively incorporated in the resource capabilities section (\S~\ref{sec:tax:sys}). However, the scope of this current review is substantially wider and more in-depth, as is seen by the rest of the article.

In summary, our survey is based on the premise that: (1) it is important to consider edge, fog and cloud resources collectively and also in contrast to each other, to leverage their mutual benefits; (2) this has to be examined from the application definition and scheduling perspective as it faces the end users and developers on these resource, rather than the service providers; and (3) programming models and scheduling techniques on individual resource layers are translatable to others, in addition to exploring approaches that cut across these layers. To this end, we offer a novel and useful review of current literature and a consequent taxonomy related to these goals.

As a result, it presents designers of scheduling algorithms for edge, fog and cloud applications with a clear set of system and application features they should consider for their target infrastructure. It also provides architects of application runtimes with the available options of scheduling algorithms that they can leverage to meet the needs of their end-users. Also, it highlights gaps in existing literature where the intersection between these resource layers have yet to be adequately addressed.

\section{Background}

In this section, we provide a background of edge and fog computing to motivate the need to consider them as first-class computing resources, while substantiating this with a conceptual taxonomy for them later in \S~\ref{sec:tax:sys}. 
We then briefly discuss prior work on Mobile Cloud Computing (MCC) (also called, Mobile Edge Computing (MEC)), 
which has generalized into edge and cloud computing. This offers a contrast from this conceptual predecessor, and 
scheduling strategies that have been attempted on it. Lastly, we offer a similar distinction from the extensive work on scheduling for High Performance Computing (HPC) resources, which is related to but has key distinctions from 
how applications are scheduled on the cloud. 
These establish clear contrasts from our effort while still offering a background on prior work on related technology domains.

\subsection{Edge and Fog Computing as an Emerging Resource Abstractions}

\label{sec:ef:abs}

\emph{Edge computing} refers to the use of thousands of computing devices such as sensors, gateways, mobile devices or embedded systems at the edge of the network, often in the context of mobile phones or the Internet of Things (IoT), for performing computation. This complements their traditional role of data collection and actuation, while the cloud is used for computation and data analytics~\cite{hu2017survey,Bonomi2014}. \emph{Fog computing}~\cite{Bonomi2014}, also known as \emph{Cloudlets}~\cite{satya:pervasive:2009}, was introduced by Satyanarayanan, et al., and popularized by Cisco as a complementary resource-rich layer that sits between the edge and the cloud~\cite{varshney2017icfec}. 
Fog provides data, compute, storage, and application services to end-users similar to cloud data centers but with lower latency and faster response as it is typically 1-hop away from the edge~\cite{bittencourt2017mobility,stojmenovic:fedcsis:2014}. 

There are many \emph{applications} that motivate and benefit from edge and fog computing~\cite{varshney2017icfec}.
        There is a global push toward Smart Cities as a manifestation of IoT. 
Given the advances in deep learning, large-scale \emph{video surveillance} has been adopted for public safety and as a proxy for ambient observations using analytics, such as to identify parking violations, and classify vehicles and people
~\cite{yi:mobidata:2015,jin2012network}. Training the neural network models is computationally costly, and the source video streams at the edge are also large in size. Accelerated fog resources can complement edge resources available for deep learning while reducing data transfer costs to the cloud.
\emph{Smart Power Grids} are another key domain in smart cities~\cite{simmhan:cise:2013,varshney2017icfec}, 
with net-connected smart meters reporting power demand at households and industries every few minutes to the utility
~\cite{amin2005toward}. 
Smart grid applications like \emph{Demand-response (DR) optimization} help shape or shift power demand 
using forecasting models on the cloud that trigger curtailment strategies on the edge when a load mismatch is detected~\cite{aman:tkde:2015,aman:smartgridcomm:2015}. 
\emph{State estimation} to determine the health of the distribution network is even more time sensitive, $\mathcal{O}(ms)$, and necessitating computing at the edge~\cite{moslehi2010reliability}. 

While edge and fog computing are still emerging technologies, this taxonomy throws more light on these resource abstractions and their effective use from an application and scheduling model, based on current literature and technology.

\subsection{Scheduling on Mobile Clouds vs. Edge Computing}
\label{sec:mcc}

Mobile Cloud Computing (MCC) (also called Mobile Edge Computing (MEC) or Mobile Clouds) is a precursor to the more general edge computing concept~\cite{dinh2013survey,sanaei2014heterogeneity,fernando2013mobile}. 
It has a restricted design, motivated by cellular phones, both smart and feature phones, running applications or ``apps''. 
MCC is concerned with strategies for offloading these applications from the mobile devices to the cloud due to the constrained computing power of the device. It involves a simple network topology, and is often limited to direct communication between one mobile (edge) device and the cloud. 

In the most common form of MCC, the coordination between the mobile device and the cloud is often on a per-application basis, i.e., each application is designed to run a part of its logic on the phone and the rest on the cloud. E.g., the cloud may be used for data persistence, to look-up information, or to perform some costly computing tasks. The interaction may be using service endpoints, and this takes the form of a client-server architecture. 

However, there exist research on more general frameworks that decide which applications or modules to offload, and when. CloneCloud~\cite{chun2011cloneCloud} partitions a mobile application and migrates it to a device clone running in the cloud to minimize the application execution time and conserve the energy of the device. ~\cite{barbera2013offload} study the trade-off between off-loading and not off-loading computation and software/data backup from mobile edge devices to the cloud, using bandwidth and energy consumption as metrics.
Others plan of such off-loading at the cellular network level for different devices to the cloud, with apps defined as workflows~\cite{deng2015computation}.

MCC usually does not cooperatively schedule apps across a group of devices, due to security concerns of phones users, and  
energy constraints of the devices. 
Also, there are typically only the mobile device and the cloud layers. 
There is some literature on 
smart phones interacting with other nearby devices for performing computations~\cite{shi2012serendipity}, while others have also proposed using cellular towers as base-stations as a fog-like layer.

We generalize this even further by considering edge, fog and cloud resource abstractions, and examining distributed scheduling across one or more of these. 
That said, MCC offers insights for such current efforts for scheduling and resource provisioning.

\subsection{Scheduling on HPC Clusters vs. Cloud Computing}
\label{sec:cloud:distinction}

Traditionally, scheduling has been an important aspect for operating systems, high performance computing (HPC) and supercomputing clusters, and computing Grids~\cite{hpc-sched,grid-sched}. 
Grid computing offers shared distributed resources for which scheduling strategies are crucial, and these have been reviewed in detail~\cite{grid-survey1,grid-survey2}. 

In contrast, cloud computing is a distributed computing capability offered by data center operators using a service-based model
~\cite{berkeley:2009}, while edge and fog computing operate on a wide-area network, and these 
affect how applications are scheduled upon them~\cite{foster:360}.

We summarize key resource distinctions of HPC that impacts application design and scheduling, and motivates the need to separately explore scheduling on edge, fog and cloud resources.

HPC centers traditionally have captive cluster infrastructure that are 
accessed  by the center users 
using a batch queue that schedules jobs based on arrival time. 
The emphasis is on how best to allocate the available resources to the waiting jobs from 10--100's of users. 
In contrast, public cloud infrastructure offer access to Virtual Machines (VM), on-demand without delay, and give the illusion of ``infinite'' resources to 1000's of users~\cite{saeid:fgcs:2013}. It also allows the number of VMs requested to be elastically scaled up and down
~\cite{imai:ucc:2012}. At the same time, each application request 10's of VMs rather than 100--1000's of cores common in HPC clusters.

Grids and HPC clusters use high-end fault-tolerant servers and high-performance networks to support Floating Point Operations per Second (FLOPS) and communications numerical applications.

Clouds on the other hand use commodity and virtualized hardware, run on Ethernet, and are not as resilient to hardware failure. They also offer VMs of different resources 
capacities 
unlike HPC nodes that are typically homogeneous. As a consequence, 
HPC clusters can host tightly-coupled large-scale applications 
with high throughput and reliability
~\cite{raicu:2008}. 
Application \emph{Makespan} 
is the primary measure of success for scheduling. Uniform nodes also limit the degrees of freedom when scheduling applications on such clusters~\cite{topcuoglu:tpds:2002}.

Clouds are popular for loosely-coupled 
applications that run from seconds to days each and have variable resource needs
~\cite{jackson:2011}. Fault-tolerance is built in software due to weaker hardware robustness, and 
applications are more delay tolerant. 
Also, VMs may have variable 
performance due to virtualization 
and multi-tenancy
~\cite{iosup:ccgrid:2011}. Factors like VM acquisition lag and diverse VM sizes add to the scheduling complexity~\cite{mao:sc:2011, 
  calheiros:tpds:2014,jackson:cloudcom:2010}.

Lastly, Grids and HPC clusters encourage full use of their high-end infrastructure by their users to amortize the high capital costs. There is little or no financial cost to the users and at best quotas are imposed for fairness. 
As a result, scheduling algorithms prioritize the performance and makespan of the applications rather than conserve the resource usage~\cite{Sandholm:2004}.

Public clouds follow a pay-as-you-go model~\cite{berkeley:2009} with resources billed only for resources acquired on-demand, 
and diverse costing models. 

These offer scheduling algorithms a different parameter space to optimize upon. 

\begin{figure}[t]
	\centering
	\includegraphics[width=0.9\columnwidth]{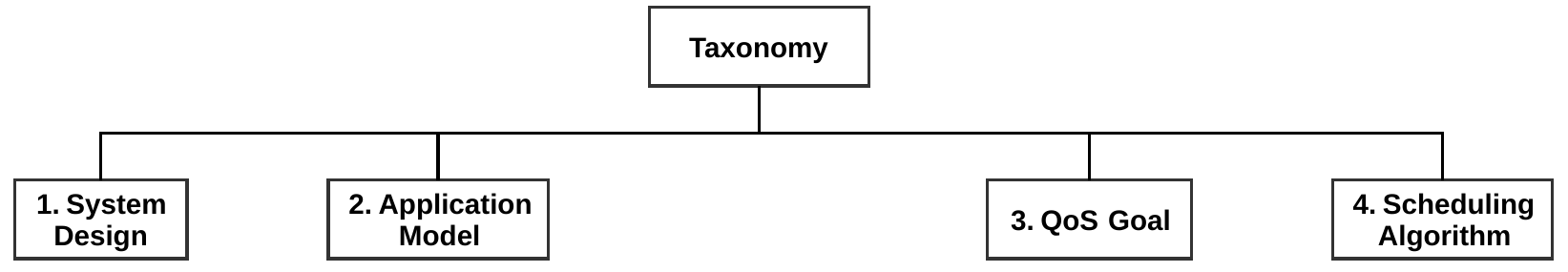}
	\caption{Taxonomy of concepts and approaches for scheduling applications on Edge, Fog and Cloud resources}
	\label{fig:taxonomy}
\end{figure}

\section{The Taxonomy}
\label{sec:taxonomy}

Our taxonomy takes a holistic view of application scheduling on edge, fog and cloud resources, and offers a classification of the \emph{conceptual model} of system models, application models, and their goals, required to design scheduling algorithms. We categorize the scheduling algorithm design themselves, and approaches used to evaluate them. Specifically, the top levels of our taxonomy are: System design (\S~\ref{sec:tax:sys}), Application Model (\S~\ref{sec:tax:app}), QoS Goal (\S~\ref{sec:tax:qos}), and Scheduling Algorithm design (\S~\ref{sec:tax:algo}), as shown in Fig.~\ref{fig:taxonomy}. A full view of the taxonomy is provided in the Appendix, in Fig.~\ref{fig:tax:full}. In the following sections, we discuss each category in detail.

\subsection{System Design}
\label{sec:tax:sys}
\begin{figure}[t]
	\centering
	\includegraphics[width=0.85\columnwidth]{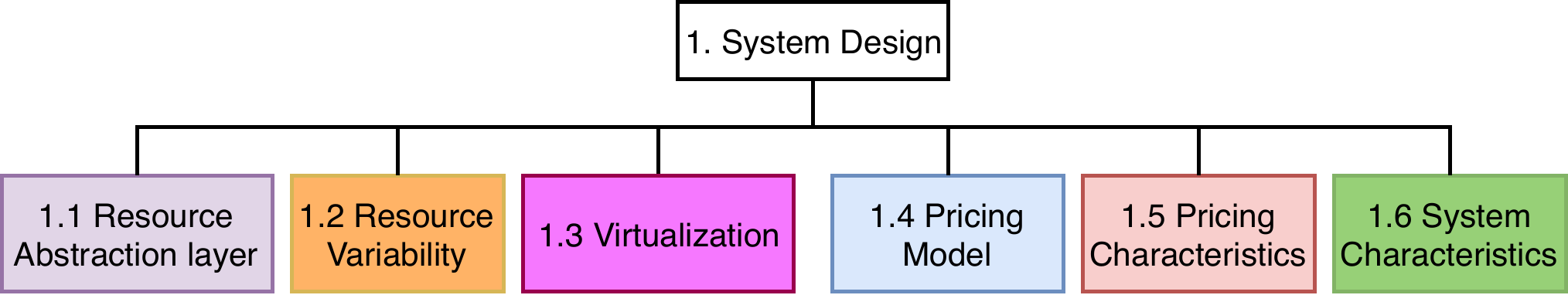}
	\caption{Taxonomy of system design}
	\label{fig:tax:sys}

\end{figure}

Edge, fog and cloud provide various types of computing resources with diverse capabilities and different pricing models. System design is concerned with the resource capacities, pricing features, and other system characteristics (Fig.~\ref{fig:tax:sys}). For the cloud layer, we base our characterization on the capabilities of popular public Infrastructure as a Service (IaaS) clouds from Amazon Web Services (AWS), Microsoft Azure and Google Cloud Engine in defining these dimensions. For the edge and fog resources, these are based on current research and early commercial offerings such as Amazon Greengrass and Azure IoT Edge. 
We distill the essential and generic capabilities of such systems, and avoid transient capabilities offered in this fast-changing landscape.

\subsubsection{Resource Abstraction Layer}

In this section, we offer a relative overview of the three resource layers on which the applications are scheduled, as highlighted in Fig~\ref{fig:pyramid:1} and examined before~\cite{varshney2017icfec}. We have already introduced these resource classes, and now we contrast their resource and performance characteristics. 

We base these on existing definitions~\cite{brogi2017qos,shi2016Edge,vaquero2014finding,mell2011nist,buyya2008market}, among many, which place fog computing as a resource layer that fits between the edge devices and the cloud data centers, with features that resemble both. 

The intrinsic distinction is the \emph{network distance} between the edge/leaf of the Internet, where edge and fog resources are present, and the core of the Internet where large cloud data centers are located. This affects the latency and available bandwidth between the different layers. This combined with where the data is generated, analytics are computed, decision signals are sent, and what QoS is required can affect the scheduling problem. 

\begin{figure}[t]
		\centering
	\subfloat[Physical presence vs access]{\includegraphics[width=0.4\columnwidth]{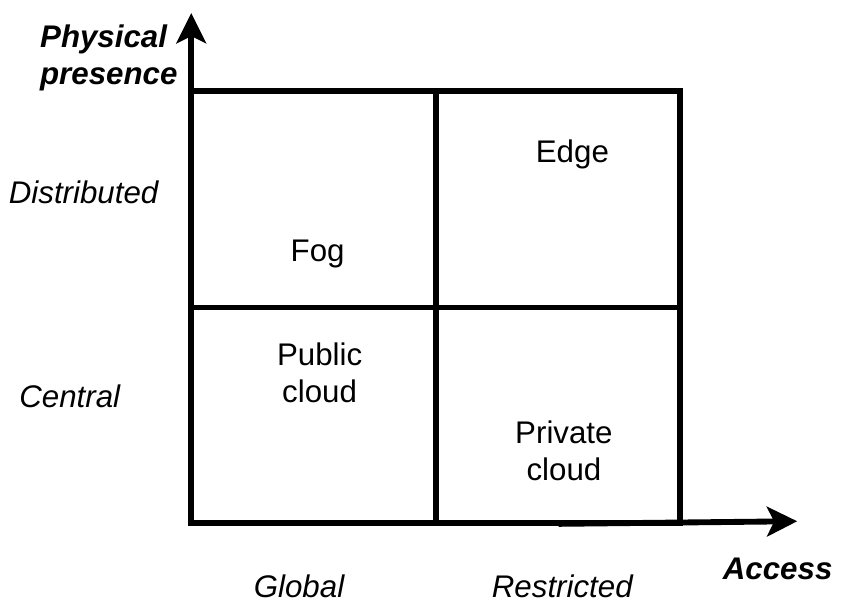}\label{fig:physical}}\qquad\qquad
			\subfloat[Physical mobility. Cloud is omitted as it is static.]{\includegraphics[width=0.45\columnwidth]{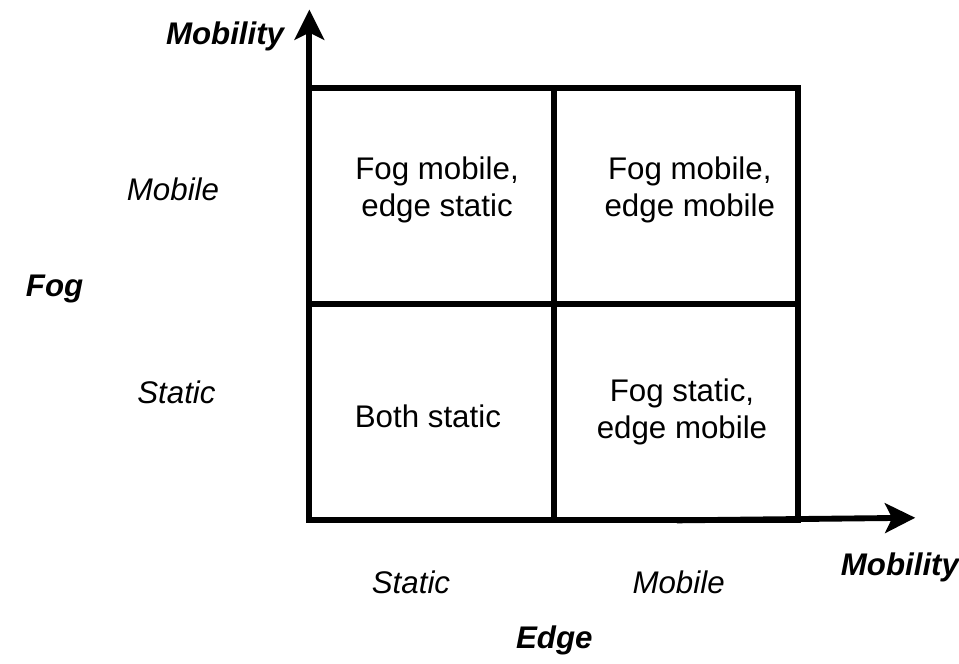}\label{fig:mobility}}
			\caption{Physical presence, access limits and mobility of Cloud, Fog and Edge resources~\cite{varshney2017icfec}}
		\end{figure}

In addition, it is also worth considering the \emph{physical distance} between the three computing paradigms, and their accessibility by clients. In Fig.~\ref{fig:physical}, four quadrants are formed from considering whether the resources within a layer are physically centralized or distributed (Y Axis), and whether their access is global or restricted (X Axis).

		Resources in a cloud data center are centrally located, but depending on whether the cloud is public or private, are available to anyone in a pay-as-you-go model, or only to users of the private corporation~\cite{luan:arxiv:2016}. That said, public cloud providers host geographically distributed data centers, sometimes several in a country or continent, while the number of large private data centers for an enterprise is more limited.

                Edge resources such as smart phones and set top boxes are distributed far and wide, but their access is restricted to individual users or managed applications~\cite{Bonomi2014,yi:mobidata:2015}. Fog resources are also physically distributed to be close to the edge, but not as dispersed. Additional specializations, on whether there is a fog for each city block, one for the whole city or other variants, depend on the business models and applications that will evolve. One also expects the fog to offer as a shared, pay-as-you-go IaaS or Platform as a Service (PaaS) model~\cite{bittencourt:pgcic:2015,yi:mobidata:2015}.
		
	    Edge and fog resources are distributed which increases the probability of \emph{attacks and failures} as discussed in \S~\ref{sec:tax:sys:char}.
		The access restrictions on private clouds and edge devices translates to a \emph{zone of trust} for applications and services hosted on them, which enables sensitive data and services to be hosted on them. Fog and public clouds, however, are designed as shared resources with \emph{multi-tenancy}, which require higher measures of security and sandboxing.
		That said, there may be fog architectures where the resources are deployed for specific applications or organization (e.g., a Smart City municipality), similar to a private cloud~\cite{luan:arxiv:2016}. Further, the fog may sit at the boundary between public and private networks, and 
                run proxy services that translate from one zone of trust to another, one service layer to another (e.g., CoAP to HTTP), or one network protocol to another (e.g., IPv6 to IPv4).

It helps to understand the impact of \emph{mobility} on these three resource layers, as illustrated in Fig.~\ref{fig:mobility}, as this impacts the communications, applications and platform design. We distinguish between mobility of the \emph{physical resource}, discussed here, and mobility of the logical applications, which we examine later.

		Cloud data centers, obviously, are not mobile though their platforms can ease the mobility of data and applications among their data centers. Spatial mobility at edge devices is frequent, though not universal, e.g., mobility is seen in ubiquitous smart phones and autonomous vehicles, while they remain static in traffic cameras and smart power meters~\cite{stojmenovic:fedcsis:2014,Loke2015TheIO}. This is more so in the context of mobile cloud computing 
                that is concerned with offloading computations from mobile edge devices like smart phones to cloud to save battery, speedup computations and for data/software backup~\cite{barbera2013offload}. 
Likewise, the fog layer can also be manifest as a static or mobile resource~\cite{luan:arxiv:2016}. A fog server can be installed at fixed sites such as a coffee shop or the airport, or on mobile vehicles such as taxi cabs or trains. 
                This mobility can cause these resources to be unavailable which is discussed later in \S~\ref{sec:tax:sys:char}.

Mobility at edge and fog layers also necessitates \emph{device discovery}, as devices join/leave/rejoin different parts of the network and resource fabric. This will on-board and make them available as part of the resource pool, and similarly remove them when they leave. The \emph{Open Fog Reference Architecture}~\cite{ofra} suggests a P2P model where a new fog node broadcasts its information to a fog cluster. Other have proposed a publish-subscribe model for edge and fog devices to announce themselves on arrival/departure~\cite{7721750}. This can be extended to a Distributed Hash Table (DHT) as well, for notifying arrival and departure, as well as for scheduling tasks~\cite{gedeon2017router}. Some also suggest a hierarchical discovery approach for edges partitioned into fog parents, and fogs themselves reporting to higher-level fogs, all of which is accumulated at a \emph{discovery server}~\cite{saurez2016incremental}. Some also use the transport-level protocols~\cite{soo2016proactive}, e.g., by having a \emph{leader device} broadcast a 802.11 WiFi beacon frame to notify spatially proximate devices that wish to join about the location of the leader to contact~\cite{rejiba2018f2c}.  
Discovery using the Software Defined Networking (SDN) layer is possible as well, as has been suggested for fog resources in a vehicular network~\cite{truong2015software}.

		As a result, depending on the mobility of the edge and fog layers, the application and platform will need to be designed based on \emph{permanent, transient, periodic, or ephemeral connectivity} between the layers and within the layers which can determine the \emph{reliability of access} to data, storage, network and computing resources.

The application definition needs to be scheduled and coordinated in order to meet various QoS goals such as latency, energy and monetary constraints. This coordination can be done using different strategies, across edge, fog and cloud resource layers. Three common \emph{orchestration models} that are relevant in such a multi-layered and distributed resource environment are \emph{centralized, hierarchical} and \emph{peer-to-peer (P2P)}. We also distinguish between \emph{scheduling decisions}, the \emph{flow of control signals} and the \emph{flow of data}, and different coordination models could be applied to these.

\emph{Centralized} orchestration has a single service, either per application or for the platform, that is located in one of the three resource layers, makes scheduling decisions, and coordinates the transfer of control signals and/or data items~\cite{dasterdi:corr:2016}. 
This is simple to design but can suffer from high latencies and transfer costs, and is a single point of failure. While this orchestrator often runs in the cloud (to coordinate across edge devices) or the edge (to interact with different cloud services), the fog layer could offer a sweet-spot for such a centralized coordinator~\cite{aazam:2014}.

A \emph{hierarchical} architecture is a generalization of the centralized model, and allows only vertical communication of data and controls to take place between adjacent layers. This is a natural fit for fog computing as it leverages both the bandwidth and latency benefits of the fog layer in accelerating these flows, as well as the compute benefits closer to the observation source~\cite{yannuzzi:camad:2014,yi:mobidata:2015,stojmenovic:fedcsis:2014,dasterdi:corr:2016,hong:mcc:2013}. Often, the cloud forms the root of this tree and is used for global data aggregation and coordination. Local data analytics is delegated to Cloudlets and further to the edge devices. This allows a federated behavior that has shown to scale.

\emph{P2P} is a form of distributed coordination that avoids a single point of failure. Here, peers in the same edge or fog layer  
can pass control and data directly among each other~\cite{vaquero2014finding}. The horizontal communication channels may initially be setup by an entity that has a global picture of the resources. This is typically done at the cloud or the fog, or one of the edge devices that serves as a leader. There are simple component-based models for composing and executing P2P applications, as well as complex ones that use Distributed Hash Table (DHT) to maintain an overlay network over peers that frequently enter and leave the system~\cite{luan:arxiv:2016,stojmenovic:2014}.

In a \emph{hybrid} model, there are no strict limitations on the flow of control or data flows, and all layers are seen as having resources of heterogeneous characteristics. While there can be interconnections among resources within each layer (cloud, fog, edge), communication can also take place vertically~\cite{bittencourt:pgcic:2015}. This can 
require more complex coordination, but can potentially improve the resilience of the application when network connectivity between specific layers is interrupted~\cite{madsen2013reliability}.

\subsubsection{Resource Capacities and Variability}

\begin{figure}[t]
	\centering
	\includegraphics[width=0.7\columnwidth]{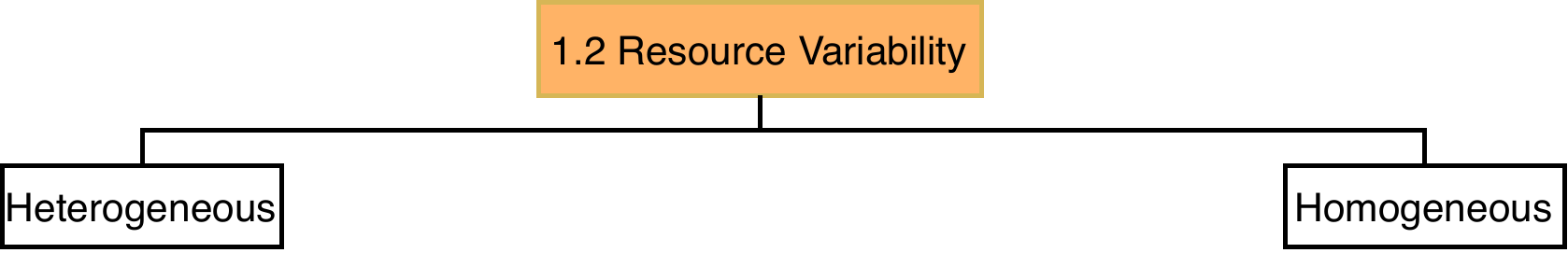}
	\caption{Taxonomy of resource variability.}
	\label{fig:tax:sys:vm-size}

\end{figure} 

\begin{figure}[t]
	\centering
	\includegraphics[width=0.75\columnwidth]{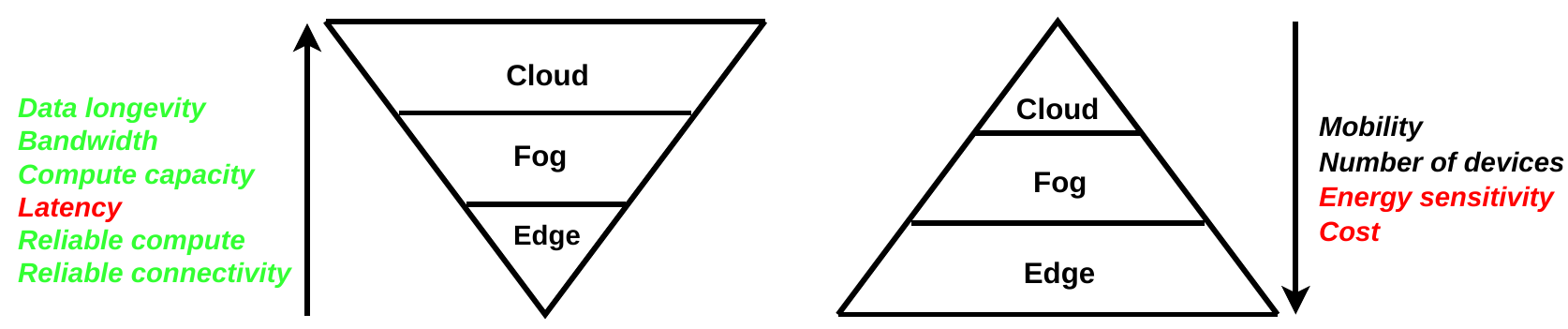}
	\caption{Resource characteristics of Cloud, Fog and Edge computing systems~\cite{varshney2017icfec}}
	\label{fig:pyramid:1}

\end{figure}     

Next, we consider the variability of the resource capacities offered on the different abstraction layers, with the shown in Fig.~\ref{fig:tax:sys:vm-size}. Schedulers often leverage the ability to acquire resources of variable types and capacities as it allows a best-fit between the application resource requirements and resources. Resources can vary in their \emph{types}, \emph{sizes} and \emph{capacities} across edge, fog and cloud layers. This is illustrated in Fig.~\ref{fig:pyramid:1} as inverted/upward pyramids that form a continuum across the layers, in increasing and decreasing order as applicable.

\ssec{Homogeneous resources}
Cloud computing resources are standardized within a service provider, and offered in different capacities based on a pay-as-you-go pricing model. 
IaaS providers offer Virtual Machines (VM) as their computing resource, characterized by their 
different compute capacities (CPU cores, clock speed, architecture generation), physical memory, and network bandwidth, with associated pricing. These VM sizes are crucial in the context of scheduling cloud applications, and one can rely on the cloud provider to offer numerous homogeneous instances of a single resource size.

Homogeneity may be possible in large scale deployments of edge and fog, as a commercial service operation or by city utilities. 
Edge devices that are part of city-scale IoT infrastructure, such as net-connected smart power meters and traffic cameras for machine-to-machine (M2M) interactions, may have uniform specifications~\cite{simmhan:cise:2013}. 
Fog resources such as Nvidia Jetson TX1 and Dell Edge Gateway can also be deployed as a standard across the city for commercial use, such as Barcelona city's ``street-side cabinets''
~\cite{yannuzzi-2017}.

          \ssec{Heterogeneous resources} 
At the cloud layer, a wide variety of VM sizes and configurations are offered, with AWS, e.g., offering $42$ different Elastic Compute Cloud (EC2) VM configurations. Besides resource capacities, these also vary in types of disks (SSD or HDD), presence of accelerators (GPUs), and higher-end architectures (DDR4 memory, latest CPU generation). 

While in the cloud, heterogeneity is a choice offered by the provider, on the edge and fog, this variability may be a necessity of the infrastructure deployment model. Heterogeneity is increased when the resources are consumer-owned instead of being available as part of commercial deployment, such as

smart phones, smart watches, Virtual Reality (VR) headsets, etc. 
Edge platforms also tend to be constrained devices, with battery or memory capacity often being the limiting factor rather than even compute capability
~\cite{dastjerdi:computer:2016,varshney2017icfec}. 
The fog layer offers compute resources with a higher capacity than the edge but to a smaller scale than clouds~\cite{stojmenovic:fedcsis:2014,Bonomi2014}. 
However, their resource capacity can vary~\cite{simmhan2018fog}, with Raspberry Pi devices at one end, to 
``micro'' or ``nano'' data-centers (MDC) on the other
~\cite{chiang-2016,lopez-2015}.
The latter allow fogs to serve as a ``reverse CDN'' to let edge devices stage data on them and eventually push them to the cloud for archival, after some pre-processing~\cite{bonomi2012Fog,satya:pervasive:2015,dasterdi:corr:2016}. 
Resources may also have variable \emph{network characteristics}~\cite{varshney2017icfec}. While the Fog is close to the edge in the network topology, 
whether it is at a 1-hop or multi-hop depends on the deployment~\cite{satya:pervasive:2009}. 
The fog is also expected to have a reliable and high bandwidth Internet link~\cite{stojmenovic:2014}, but its connectivity to the edge may be less robust due to the use of wireless links for the last mile~\cite{luan:arxiv:2016}. 
	
	However, such diversity explores the dimensions that the scheduling algorithm has to examine. 
        To mitigate this, some algorithms assume a uniform, homogeneous resource capacity (and price) for simplicity, which 
        is more feasible 
        on cloud resources~\cite{malawski:sc:2012}. Algorithms may also limit themselves to leveraging just the compute diversity (e.g., different VM or container sizes)
~\cite{calheiros:tpds:2014,chu:ipdps:2014,poola:iccs:2014,hong:mcc:2013}.

\begin{figure}[t]
	\centering
	\includegraphics[width=0.75\columnwidth]{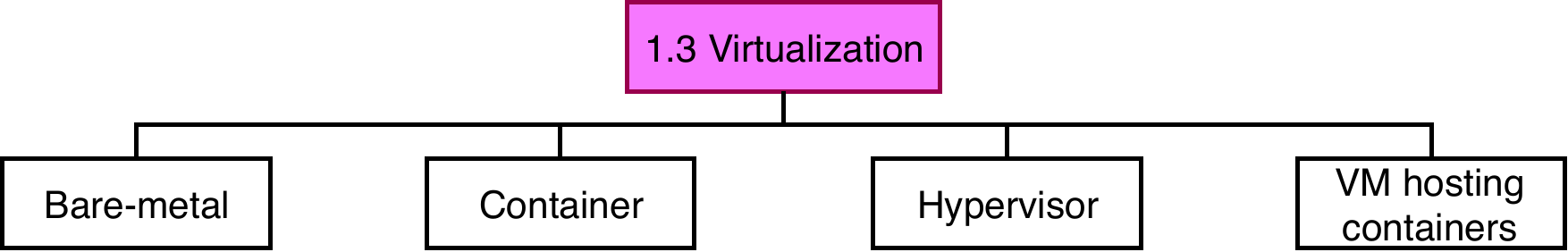}
	\caption{Taxonomy of virtualization of resources}
	\label{fig:virtualization}

\end{figure} 

\subsubsection{Virtualization}
Various types of virtualization (or lack of it) is possible within the different resource types, as shown by the taxonomy in Fig.~\ref{fig:virtualization}. Clouds expose every resource ``as a Service'' using \emph{fabric software} for infrastructure management, and this in part is a reason for its success~\cite{bittencourt:pgcic:2015,stojmenovic:fedcsis:2014,varshney2017icfec}. 
Virtualization using \emph{hypervisors} offers two key benefits: (1) custom OS and software environments, and (2) sandboxing of VMs from each other using hardware-level support. The former allows user applications and Big Data platforms 
to work equally well on different clouds. 
They latter provides dependable resource allocation and security. 

Resource-rich fog devices may be able to support hypervisors, 
and this can hasten their adoption for IaaS similar to clouds~\cite{bittencourt:pgcic:2015,varshney2017icfec}. On the other hand, resource-light fogs can use \emph{containers} like \texttt{lxc} and Docker which offer users the control over the software environment and limited resource sandboxing~\cite{yannuzzi:camad:2014,bittencourt:pgcic:2015}. However, security within multi-tenant containers on a host is still a concern.   
That said, containers have limited memory overheads and rapid bootup time compared to VMs, and this makes them preferred even for application management in the cloud. \emph{Containers launched within VMs} is a growing practice, with VMs offering the resource and security sandboxing and billing units, while containers (potentially 10's in a VM) allow application environment management.

Application management on the edge is a challenge, and done in either \emph{ad hoc} or tightly coupled to a platform like Android using sandboxed ``apps''~\cite{varshney2017icfec}. 
Lastly, even these light-weight wrappers may not be viable on severely constrained edge platforms. Hence, application may run on \emph{bare-metal}, or within a platform (rather than infrastructure) as a service.  

There are also \emph{software fabrics} that help manage such virtualized or containerized computing infrastructure, on the cloud, fog and edge. These serve as a form of distributed OS. While public clouds like Amazon, Microsoft and Google run their proprietary IaaS fabrics, \emph{OpenStack}~\cite{openstack} offers a full-suite of compute, data and network virtualization services for private clouds. \emph{Kubernetes} from Google focuses on container and compute management on large-scale clusters~\cite{kubernetes}. Both have also been extended to operate on fog and/or edge computing devices. While some have tried to use OpenStack as is on fog resources~\cite{yi2015fog}, others extend its features to be customized for specific limitations of edge and fog resources. Lebre, et al.~\cite{lebre2017revising} propose a decentralized P2P variant of OpenStack's Nova compute service to enable wide-area computing resources, while Chang, et al.~\cite{chang2014bringing} extend the Quantum virtual network networking to support devices present on a Network Address Translation (NAT) network. OpenStack is also natively working on porting their capabilities to edge, fog and Micro Data Center (MDC)~\cite{openstack-fog}.
Similarly, Kubernetes has been used for compute containers on Raspberry Pi-class edge devices~\cite{tsai2017distributed}, and used to deploy software on the fly on fog resources~\cite{hong2016dynamic}. Besides these, there are also open-source fabrics specialized for edge and fog, such as Eclipse Kura and EdgeXFoundry~\cite{kura,2017feasibility, edgex} that are evolving.

\subsubsection{Pricing Model}
\begin{figure}[t]
		\centering
		\includegraphics[width=0.8\columnwidth]{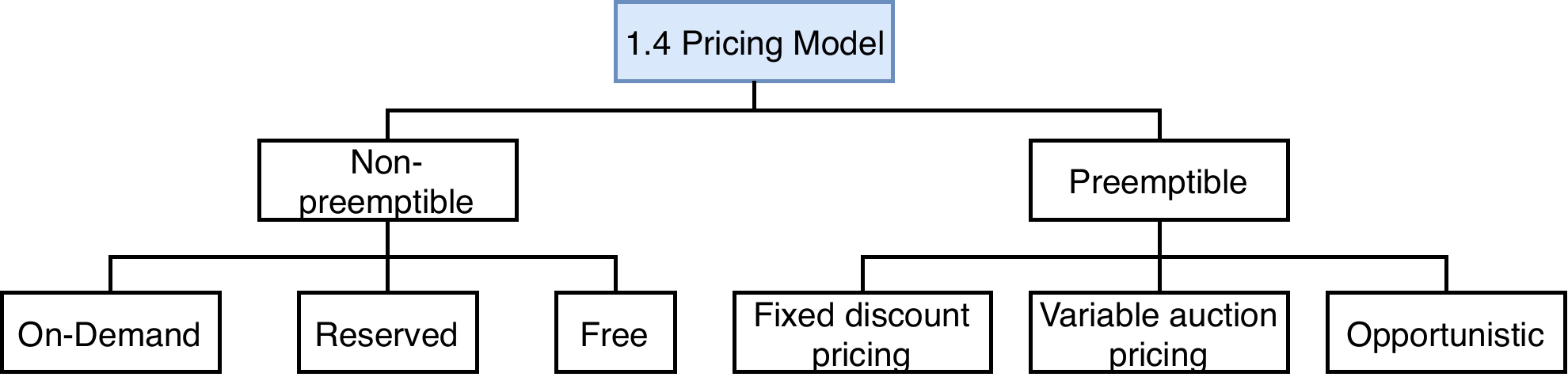}
		\caption{Taxonomy of pricing model of resources.}
		\label{fig:tax:sys:vm-price}

	\end{figure}

The monetary cost for accessing edge, fog and cloud resources is variable. While clouds offer a mature pay-as-you-go pricing model, edge and fog resources as yet have evolving business models. 
Typically, resource pricing is proportional to its compute capacity on a given resource abstraction. However, there are also pricing differences due to the variable access guarantees that are provided. One can broadly categorize the pricing model based on resources that are non-preemptible and those that pre-emptible (Fig.~\ref{fig:tax:sys:vm-price}). In the former, resources once provided are retained (and billed) until the application relinquishes them. In the latter, resources may be taken back by the service provider at any time, and the application is optionally compensated monetarily.

\ssec{Non-preemptible} These are the most common form of cloud resources where VMs have an associated fixed price per unit billing time based on their compute capacity. Clouds further distinguish this between \emph{on-demand VMs} which are acquired and released by the application flexibly, based on their current compute needs, and their billing intervals are as low as 1~second. \emph{Reserved VMs} are those that are acquired in bulk for longer periods of time (e.g., 1 month) at a cheaper unit rate, but billed in full irrespective of their usage. These are well-suited for users having a predictable long-term workload. The diversity in cloud VM sizes also implies an associated diversity in the pricing of the resources. Variable sized on-demand VMs allow scheduling algorithms to make smart choices in trading off price to performance for their application~\cite{zhou:tcc:2014,saeid:fgcs:2013,malawski:sc:2012,mao:sc:2011,calheiros:tpds:2014}.

Commercial fog providers may use \emph{consumption-based} pricing where the users are billed as per their usage, or \emph{subscription based} pricing where the users pay a monthly fixed price and can use the fog resources network-wide, without being pinned to a particular instance~\cite{bittencourt:pgcic:2015,varshney2017icfec}. These are similar to the on-demand and reserved models in clouds. The infancy of fog deployments and their potential providers 
has implications on the operational costs as well~\cite{simmhan2018fog,vaquero2014finding,yi:mobidata:2015}. 
Alternatively, smart cities may make them available as a utility service for free or based on payment~\cite{yannuzzi-2017,luan:arxiv:2016,simmhan2018fog}. 
This may also extend to edge devices that are part of the city's deployment.
	
\ssec{Pre-emptible} Cloud providers may have spare capacity in their data center which are rented at a lower price than their on-demand counterparts, even $10\times$ cheaper~\cite{agmon2013deconstructing}. This allows providers to increase the utilization and revenue of their data centers to offset the static operational overheads. While these VMs offer the same performance as a similarly sized on-demand VM, they are not guaranteed to be available for the user's required duration and may be revoked by the provider. This requires scheduling algorithms to actively manage application checkpointing for reliability. 

There are two models of such VMs available commercially. Amazon's Spot VMs use an \emph{auction-based model} that considers the highest price bids per billing interval for a VM size, 
with prices vary even within minutes. 
Amazon has recently started providing a \emph{$2$~min warning} when spot VMs are going to be revoked. Schedulers using such VMs must be aware of the current spot price and these revocation notices, but can in turn reduce application execution costs by over $80\%$~\cite{agmon2013deconstructing,chohan:usenix:2010,kushwaha:ccem:2014,voorsluys:aina:2012,subramanya2015spoton}. 
Google and Microsoft offer \emph{pre-emptible VMs} which have a \emph{fixed discounted price} that is 
significantly cheaper than their on-demand equivalents. This makes scheduling decisions easier than Amazon's spot VMs with variable pricing. However, such VMs are currently limited to being used for a maximum of 24~hours. Google gives a $30~sec$ warning before such VM are pre-empted.
	
	Most edge devices and many non-commercial fog deployments are unreliable and transient, due to mobility, device uptime and network connectivity issues. As a result, these resources have similarities to pre-emptible cloud resources where availability is not guaranteed. Here, the edge and fog resources are available \emph{opportunistically} and often for free, with no assurance that one can acquire them at a given time, or retain them for the required period. Applications may utilize the resources while they are in proximity but lose all progress if they go out of range or due to network unavailability. Mobile edge devices like smartphones are one such example 
of opportunistic computation, and schedulers like Serendipity~\cite{shi2012serendipity} 
offload tasks to neighboring edge devices to minimize the execution time and save energy.

\ssec{Hybrid} Scheduling strategies may take advantage of a mix of resources with different pricing schemes. 
	One hybrid approach is to use captive resource capacity, such as reserved VMs or private clouds/clusters which are already paid for, along with on-demand resources that are pay-as-you-go. 
Here, there are two possibilities: ``Cloud-bursting'' or ``Cloud-firsting''. In the former, the scheduler gives priority to maximizing the use of the free captive resources before moving to on-demand (cloud) resources~\cite{bittencourt:jisa:2011,guo:toit:2014,dornemann:ccgrid:2009}. In cloud-firsting, the on-demand (cloud) resources are used by default and the limited captive resources used for instantaneous capacity access by the scheduler~\cite{chu:ipdps:2014}.

A second approach is to use a mix of on-demand and pre-emptible resources to reduce the cost of execution. \cite{chohan:usenix:2010} use spot instances along with on-demand instances to speed up the MapReduce jobs on the cloud, while others propose strategies to manage the job's life-cycle on spot, on-demand and captive resources~\cite{chu:ipdps:2014,poola:acm:2016,subramanya2015spoton}.

A third approach is to use resources from different resource abstraction layers that may have variable prices. While cloud and fog resources are reliable and available, they have higher costs as well. Opportunistic edge devices may be free but unreliable. So, when end users want the benefit of both cost reduction and reliability, scheduling algorithms may need to use more than one abstraction layer. 

Further, any free edge or fog devices may also play the role of captive or pre-emptible resource, and mimic the first two approaches. 
E.g., MCC uses unreliable but cheap mobile devices along with reliable clouds for offload its computations/software/data when the edge runs low on compute, storage or battery~\cite{barbera2013offload,miettinen2010energy}.

\begin{figure}[t]
	\centering
	\includegraphics[width=0.7\columnwidth]{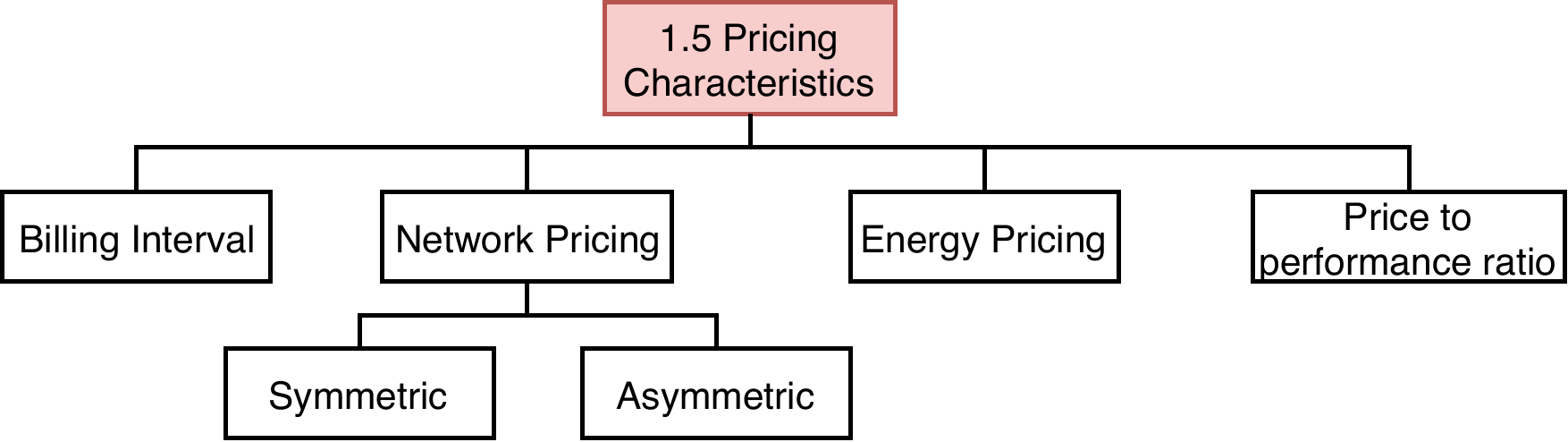}
	\caption{Taxonomy of pricing characteristics of resources.}
	\label{fig:tax:sys:price-char}

\end{figure}
\subsubsection{Pricing Characteristics}
There are other pricing characteristics of resources besides the pricing model that should be considered, as shown in Fig.~\ref{fig:tax:sys:price-char}.

\ssec{Billing Interval} In a pay-as-you-go model, users are charged per billing interval for which they use a resource. 
          Cloud providers such as Amazon used to have a billing granularity of $60$~mins, where each whole or partial VM hour was billed as a full hour. But this has drop down to 
per-second billing over the last few years, sometimes  
with some minimum time, such as $10$~mins, that is charged. This has a consequence on the sizes of applications that are scheduled on VMs, the temporal granularity of control required by the scheduler, and the price paid. E.g., in \cite{saeid:fgcs:2013}, two different time intervals of $1$~hour and $5$~mins are considered to evaluate the cost of scheduling, and as expected the normalized cost is lower when considering the shorter billing interval.

	Billing models for edge and fog resources are still evolutionary and there are not many commercial deployments that exist. They may under go a similar pricing evolution as cloud, over time.

\ssec{Network Pricing} While the pricing models above focus on compute resources, the cost of network bandwidth into and out of a resource layer may be charged, say, for GigaBytes of data transferred. This can be between Edge-Fog, Fog-Cloud, Edge-Cloud, or between resources in the same layer. 

	Network pricing can either be \textit{symmetric} or \textit{asymmetric}. In the former, the price for both moving data in and out of a resource layer is charged equally, while in the latter the bandwidth charges are different based on the direction. Often in cloud data centers, data-in and intra-data center bandwidth are kept free to encourage hosting data in the cloud. These impact the costs for moving input/output data to/from the application on the cloud and the user's machine, as well as decisions regarding using captive off-cloud resources that require migrating application state over the network. 
	
If edge and fog are deployed by the same provider or are a part of the same private network, such as a WiFi access point, then there may be no pricing costs for the network usage~\cite{simmhan2018fog}. The data transfer within a layer and between these layers will be free. However, if edge and fog are part of different networks or the capacity of a constrained network is being saturated, then the data transfers may be chargeable. Also, connectivity among edge devices on different networks may be through gateways and proxies present on the fog or the cloud, to account for firewalls and network visibility. There can be additional charges for such redirection. 
	Network pricing between edge or fog and the cloud layer is dependent on the deployment scenario. 
        E.g., 
edge/fog can be connected to the cloud with a broadband connection or 4G network which the local ISP may charge for. 
	
\ssec{Energy pricing} 
          The \emph{energy profile} can influence the capability and availability of some resources~\cite{varshney2017icfec}. Cloud data centers reduce their energy footprint, but to limit operational costs~\cite{bittencourt:pgcic:2015}. The fog is expected to run off grid power and, like the cloud, be energy conscious to lower pricing~\cite{vaquero2014finding,Loke2015TheIO,dasterdi:corr:2016}. But there may be remote places where the fog runs on renewables like solar, when energy-aware usage is a key goal. Edge devices are often concerned with battery life, and the choice of using specific edge features can depend on the current battery level~\cite{mishra:iot:2015}. While resource providers usually include energy costs as part of the operational cost of a resource when pricing it, it may be possible to bill the energy costs separately based on the power consumed by an application. Alternatively, energy usage may be an application constraint or optimization specification when they run on edge and fog resources~\cite{dasterdi:corr:2016,Bonomi2014}.

\ssec{Price to performance ratio} Trade-off between resource performance and its price is important while selecting a resource. 
        Clouds leverage economies of scale and usually have the lowest operational cost per resource unit~\cite{varshney2017icfec}. The elastic nature of VMs means that cloud providers attempt to raise prices linearly with the VM size, but the performance of larger VMs may be super-linear due to reduced network latency and resource contention on the same host. Scheduling algorithms can exploit these size differences~\cite{bots-on-ondemand-vm-2},  
normalize price to performance~\cite{sakellariou:coregrid:2007}, or 
leverage pricing arbitrage on spot VMs~\cite{chu:ipdps:2014}. 
Prices of edge and fog resources are dependent on the deployment scenario. The economies of scales will come into play if fog 
deployments are done at large scales. Consumer edge devices have lower capital costs due to their mass production, and zero operational costs if maintained by the user. However, 
fog/edge resources in the field, such as for IoT, may have higher operational costs due to maintenance and hence higher price to performance. 

\subsubsection{Other System Characteristics}\label{sec:tax:sys:char}
\begin{figure}[t]
	\centering
	\includegraphics[width=0.85\columnwidth]{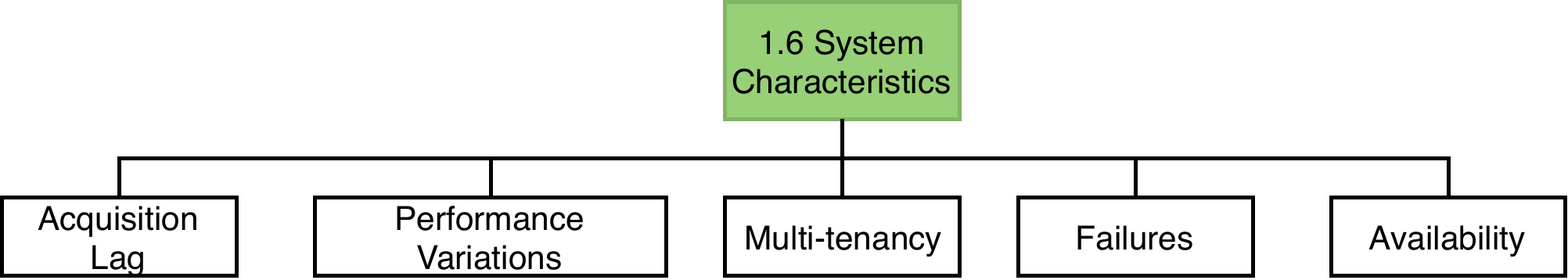}
	\caption{Taxonomy of other system characteristics of resources
	}
	\label{fig:tax:sys:vm-char}

\end{figure}

Besides resource pricing, types and sizes, several other resource characteristics impact the design of scheduling algorithms (Fig.~\ref{fig:tax:sys:vm-char}), as we discuss below.

\ssec{Acquisition lag} It might take some time for a resource to be ready for use after it is requested by a user. This delay varies with the type of resource requested and can affect the performance adversely if the scheduling algorithm frequently instantiates and switches between resources~\cite{malawski:sc:2012}.

	Cloud VMs take 10's of seconds or minutes to be provisioned, booted up, and ready for use by the end-user, and this can 
vary with the number of VMs requested~\cite{iosup:ccgrid:2011}. 
Cloud scheduling algorithms may explicitly consider this lag in their planning
~\cite{chu:ipdps:2014,mao:sc:2011}.

Edge or fog resources that run on bare-metal or on containers avoid the bootup time of hypervisor-based VMs, but may not always have adequate on-demand capacity.  
This can cause queuing delays which add to the acquisition time. Some scheduling algorithms may be able to plan deferred acquisition (or advanced reservation) where a slot is assigned for the execution of a task on a resource at some later point of time, and is available for the task with no lag at that time.

\ssec{Performance variations} Virtualized resources may not offer deterministic performance in terms of execution and data transfer time. These variations are often due to multi-tenancy where VMs collocated on the same host compete for resources or due to fabric management overheads~\cite{iosup:ccgrid:2011}. This can occur in containers too as the resource sandboxing is done by the OS and may not be as effective as hardware virtualization~\cite{violet}. These can cause the expected and observed makespan of the workflow to diverge by as much as $30\%$, and hence affect the performance of static schedules~\cite{jackson:cloudcom:2010}. 

	Some scheduling algorithms consider performance variations as a first-class characteristic when allocating resources on the cloud~\cite{calheiros:tpds:2014,rodriguez:tcc:2014,poola:iccs:2014,chen:2015,kumbhare:ccgrid:2014}. 

\ssec{Multi-tenancy}
Multi-tenancy allows different users to run their applications on the same host resource~\cite{simmhan2018fog}. The tenants may be separated from each other by VMs, containers, platforms, or not at all. Besides causing performance variations as mentioned above, 
these also impact the 
security and privacy of the applications and the data they process
~\cite{vaquero2014finding,dastjerdi:computer:2016}. 
Also, edge and fog devices may not be physically secured like a cloud data center, adding to the attack surface
~\cite{chiang-2016}. Containerization offers less sandboxing between applications than virtualization, and data and applications in the fog and edge operate within a mix of trusted and untrusted zones~\cite{lopez-2015,Bonomi2014}. This can affect the scheduling of sensitive tasks in an application to be limited to trusted resources.

\ssec{Failures}\label{failures} Cloud resources are prone to occasional failures, which may happen due to disk and memory module failures, transient errors in network, cloud fabric and hypervisor failures, and power issues~\cite{Vishwanath:socc:2010}. These failures are rare, with commercial providers promising a monthly uptime of at least 99.95\%. However, such infrequent failures can still affect the execution of mission-critical applications adversely, with some algorithms addressing such situations. Clouds being more centralized are single points of failures, but offer redundancy zones and alternate data centers. 

The wide area distributed nature of edge and fog resources increases their failure surface further~\cite{varshney2017icfec}. There is a higher chance of an edge device or fog server failing, their battery draining, or their network link dropping, and resiliency may need to be built into the scheduling strategy~\cite{madsen2013reliability}.

\ssec{Availability}\label{availability} Immediate availability of resources is a key feature of public clouds. While on-demand VMs have seemingly infinite availability, these are in practice limited to about $1000$~VMs per customer~\footnote{https://aws.amazon.com/ec2/faqs/\#How\_many\_instances\_can\_I\_run\_in\_Amazon\_EC2},\footnote{https://azure.microsoft.com/en-in/documentation/articles/azure-subscription-service-limits}. However, during periods of high demand, it is possible that a specific type of VM instance in a specific data center may not be available. 
Availability for pre-emptible VMs is naturally intermittent, while reserved instances offer guaranteed availability.

Edge and fog resources may become unavailable due to transient network failures or a discharged battery, but be back online eventually (as opposed to a failure)~\cite{varshney2017icfec,shi2012serendipity}. 
This can cause a transient loss of access to data or compute on the edge or fog. Application running across all three resources may be affected by the weakest link.

\subsection{Application Model}
\label{sec:tax:app}

\begin{figure}[t]
	\centering
	\includegraphics[width=0.9\columnwidth]{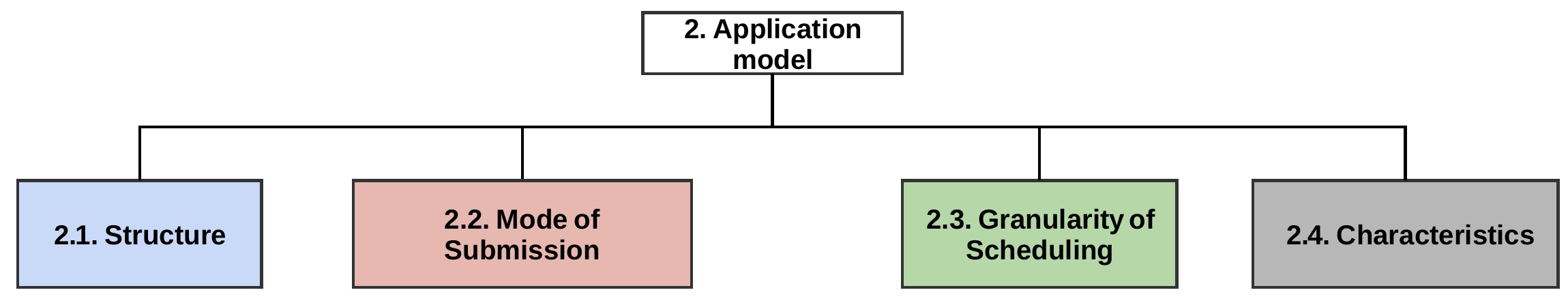}
	\caption{Taxonomy of application model}
	\label{fig:tax:app}

\end{figure}

A \emph{scheduling unit} is the unit of application submitted by the user for resource allocation on the abstraction layers, along with QoS requirements. The application model we discuss here defines the constituent members of this scheduling unit, how it arrives, and when it is scheduled, as categorized in Fig.~\ref{fig:tax:app}. The scheduling algorithm may itself use a coarser or a finer granularity of scheduling. Also, the scheduling unit can have specific characteristics which help decide the resource mapping and techniques for fault tolerance.

\begin{figure}[t]
	\centering
	\includegraphics[width=0.9\columnwidth]{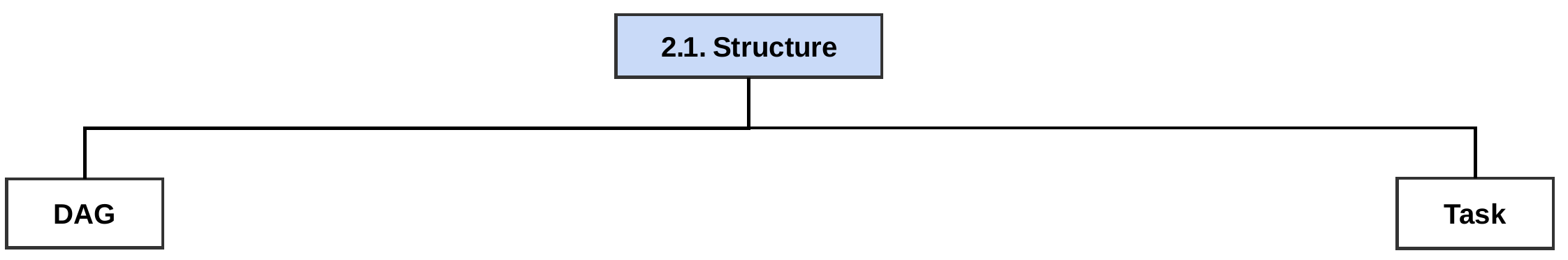}
	\caption{Taxonomy of structure of scheduling unit}
	\label{fig:tax:app:unit}

\end{figure}

\subsubsection{Structure of Scheduling Unit} 
The execution environment offered by edge, fog and cloud is inherently distributed, and the application space is vast~\cite{varshney2017icfec}. There are many application definitions in use, such as control flows, data flows, and event-driven models~\cite{simmhan2008}.  Latency sensitive applications may prefer an event-driven model that reacts rapidly to changing situations~\cite{yi:mobidata:2015}. 
Events streams are also light-weight~\cite{stojmenovic:fedcsis:2014}. An application unit can have different
structures (Fig.~\ref{fig:tax:app:unit}).

\begin{figure*}
	\centering
	\subfloat[A sample workflow DAG]{\qquad\qquad\includegraphics[height=0.15\textheight]{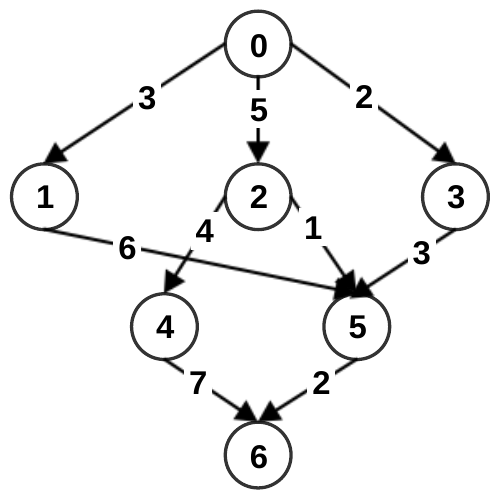}\label{fig:DAG}\qquad\qquad}~~
	\subfloat[A sample Bag of Tasks with 35 tasks]{\qquad\qquad\includegraphics[height=0.12\textheight]{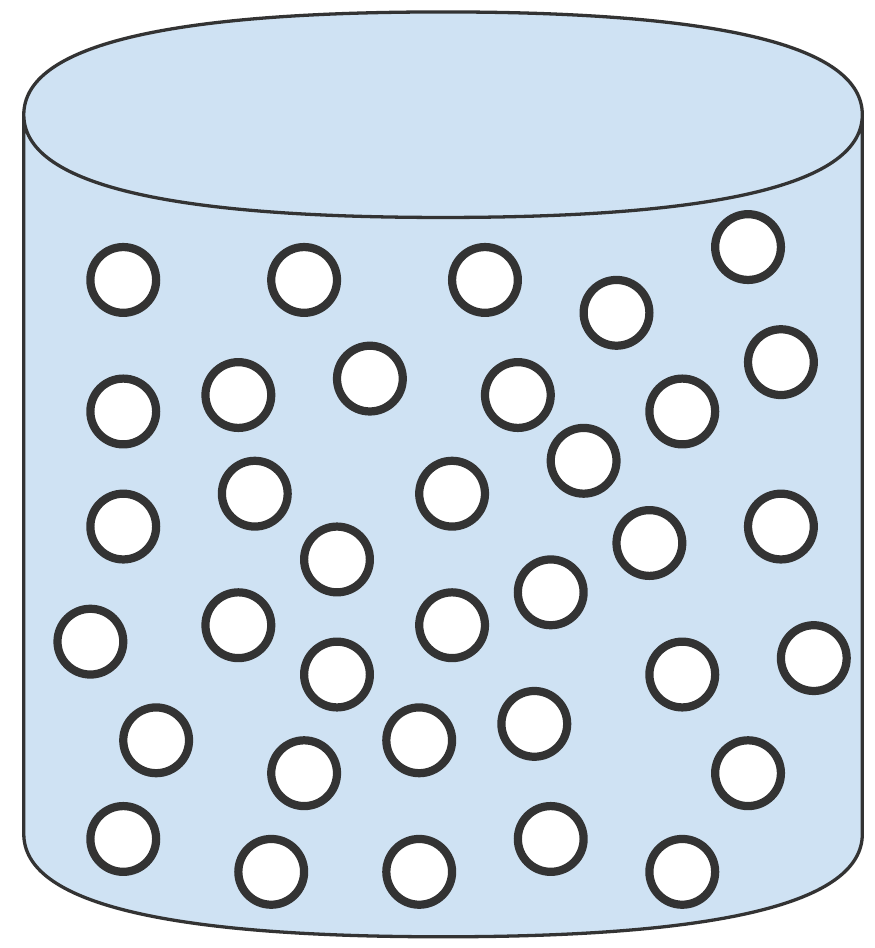}\label{fig:bot}\qquad\qquad}\\
	\subfloat[A sample Bag of DAGs with 5 DAGs]{\qquad\qquad\includegraphics[height=0.12\textheight]{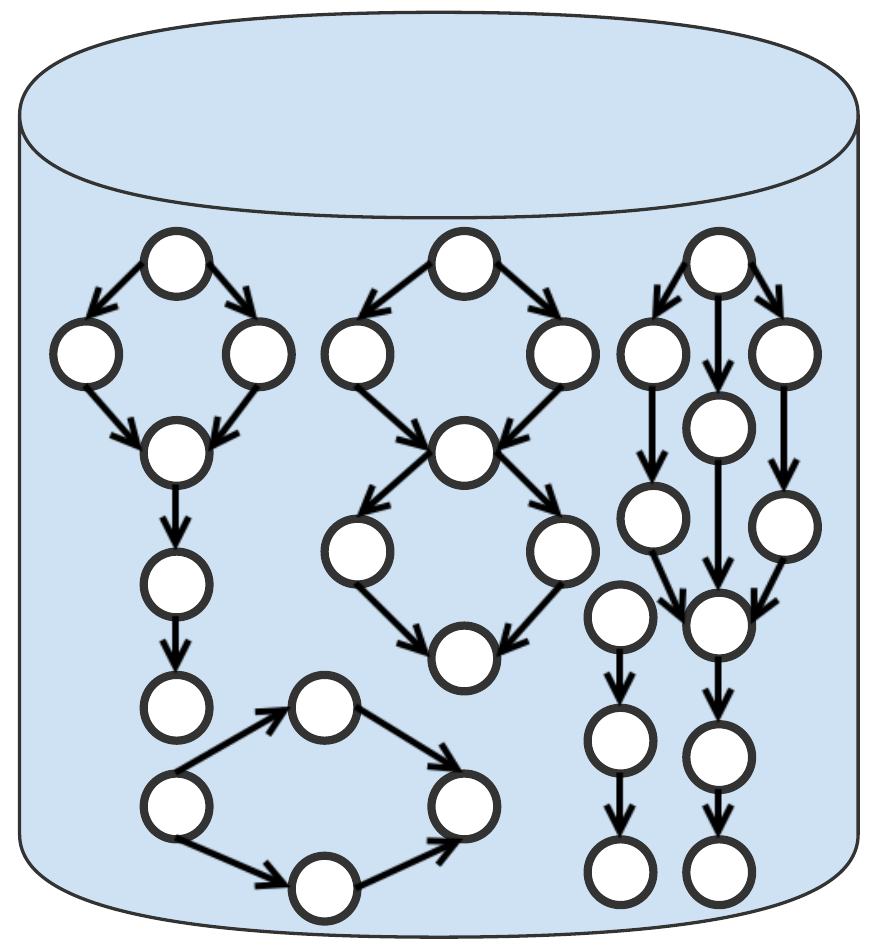}\label{fig:bow}\qquad\qquad}~~
	\subfloat[A sample Bag of Tasks and DAGs]{\qquad\qquad\includegraphics[height=0.12\textheight]{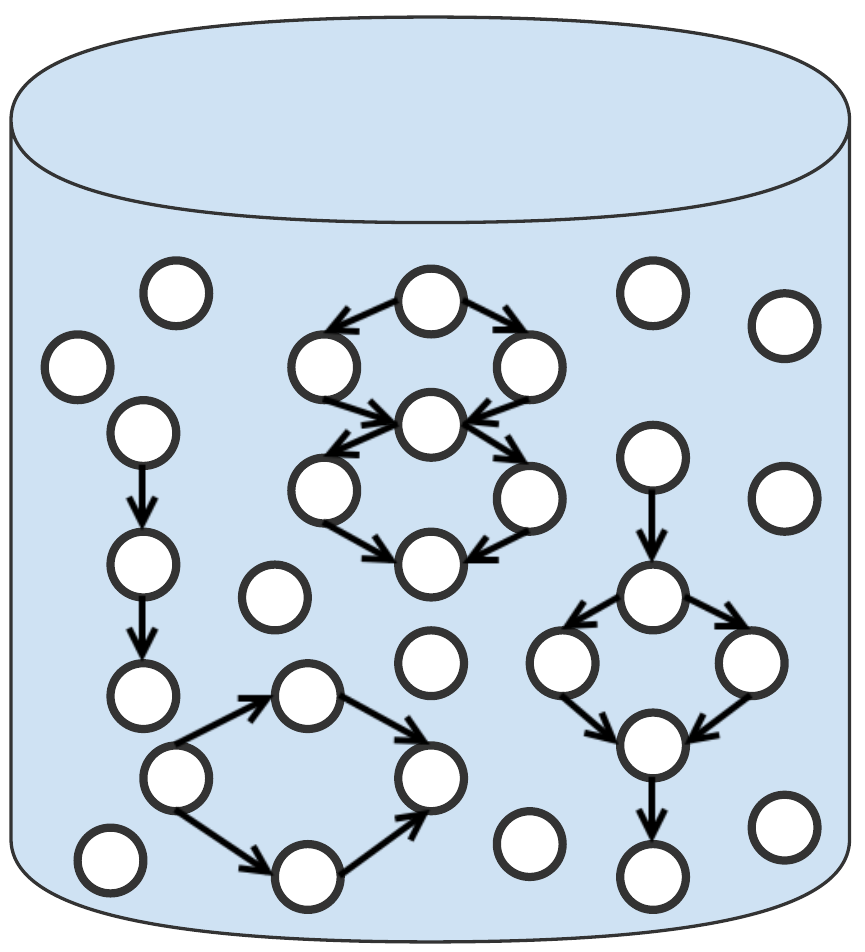}\label{fig:botw}\qquad\qquad}
	\caption{Types of scheduling units. Fig.~\ref{fig:bot},~\ref{fig:bow} are examples of homogeneous scheduling units. Fig.~\ref{fig:botw} is an example of heterogeneous scheduling units
}
	\label{fig:bags}

\end{figure*}

\ssec{DAG} Workflows, represented as a \textit{Directed acyclic graph (DAG)}, are popular for capturing flow dependencies in complex distributed applications, and are widely used as a scheduling unit provided by the user for cloud, fog and edge computing~\cite{shi2012serendipity}.
A workflow DAG is a graph \textit{G=(T,E)}, where $T$ is the set of task vertices, each of which form an \emph{atomic unit} of scheduling, and $E$ is a set of control or data dependency edges between tasks. 
A single workflow can consist of one, a few, or thousands of tasks. 
Fig.\ref{fig:DAG} shows a sample workflow with seven tasks and nine dependency edges. The number on each edge indicates the data size between the corresponding tasks, say, in MegaBytes. The dependencies can be used to identify the order of execution of the tasks. 

Workflows also offer additional information to guide their scheduling, such as the \emph{execution time} for each task on a standard resource or on each resource size~\cite{calheiros:tpds:2014,saeid:fgcs:2013}, 
or the \emph{number of instructions} required by that task~\cite{poola:iccs:2014}. 
The edges may be annotated with the \emph{size of the data} transferred between the tasks to account for network transfer time and cost~\cite{sakellariou:coregrid:2007,topcuoglu:tpds:2002,rodriguez:tcc:2014}.
The data flowing between tasks may be based on \emph{streams, micro-batches or files}. The ability to define specialized data structures, compression and transport mechanisms for distinct stream types such as video may be necessary too.
\emph{State} 
associated with a task offers a context for execution, and needs to migrate across resources~\cite{bittencourt:pgcic:2015,varshney2017icfec}. Tasks may also use \emph{location-awareness} or resource context in determining actions
~\cite{Loke2015TheIO}.

\ssec{Task} An application may be monolithic, specified as a single task. Such as task degenerates to a singleton DAG. There are many application scheduling algorithms that limit themselves to scheduling just tasks rather than DAGs~\cite{chu:ipdps:2014,varshney:tpds:2018,voorsluys:aina:2012}. For the purposes of the scheduler, a task structure is an opaque logic unit that is the smallest atomic unit of scheduling with no other task dependencies.

\begin{figure}[t]
	\centering
	\includegraphics[width=0.9\columnwidth]{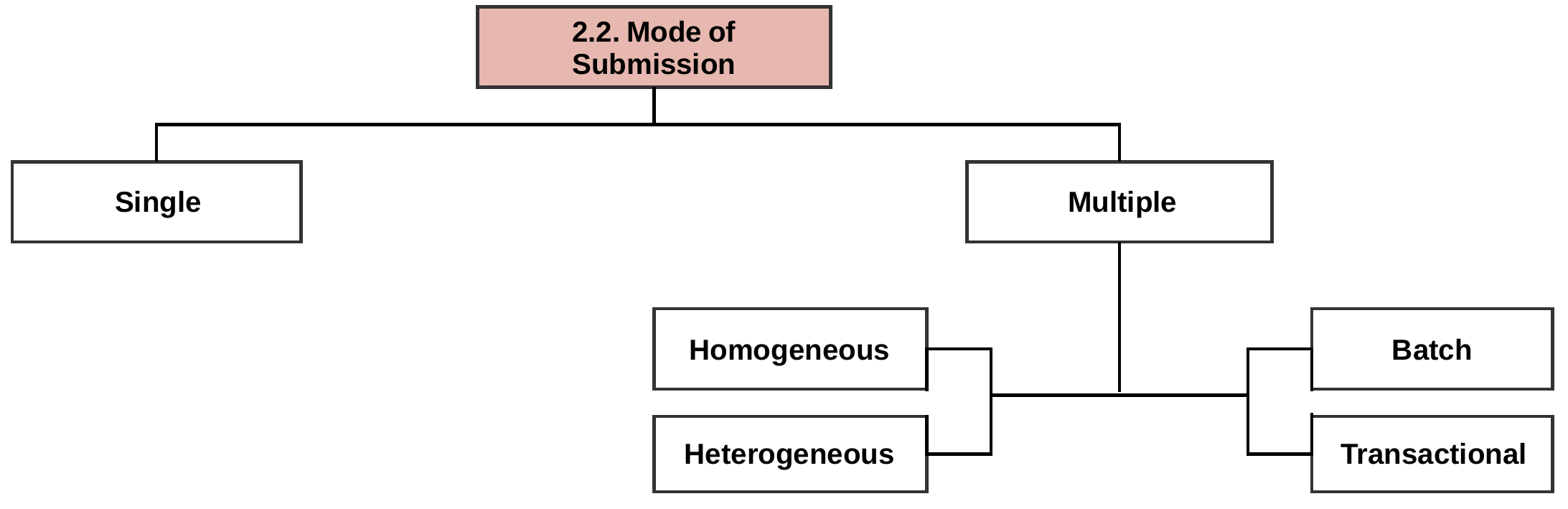}
	\caption{Taxonomy of number of scheduling units and their modes of submission}
	\label{fig:tax:app:mode:categories}

\end{figure}

\subsubsection{Mode of Submission}
Besides the unit of scheduling provided by the user, schedulers themselves may schedule individual or multiple units at a time, as shown in Fig.~\ref{fig:tax:app:mode:categories}.
A user may provide a \emph{single} scheduling unit (e.g., DAG, task), whose QoS requirements are independently specified and the scheduler schedules just this single unit. On the other hand, users may provide a \emph{set or series} of units, with associated requirements, and the scheduler needs to meet the QoS across these \emph{multiple} scheduling units. These can further be classified as \emph{homogeneous}, where all the units have the same structure, as illustrated in Figs.~\ref{fig:bot} and \ref{fig:bow}, or \emph{heterogeneous} where units can have a mix of DAGs and tasks, as seen in Fig.~\ref{fig:botw}. This distinction helps in generating more efficient schedules. Heterogeneous scheduling units, while intuitive, are less common in existing literature. \cite{wu:2013}, for example, consider both tasks of DAGs and individual tasks while generating mapping between the tasks and VMs.

\begin{figure*}
	\centering
	\subfloat[Batch mode of submission]{
		~\includegraphics[width=0.7\textwidth]{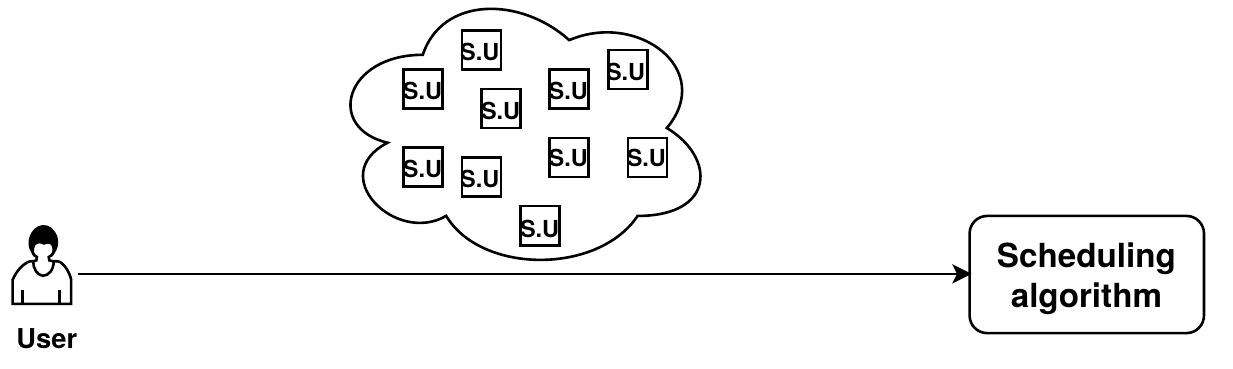}
		\label{fig:tax:app:mode:batch}
	}\\
	\subfloat[Transactional mode of submission]{
	    \includegraphics[width=0.7\textwidth]{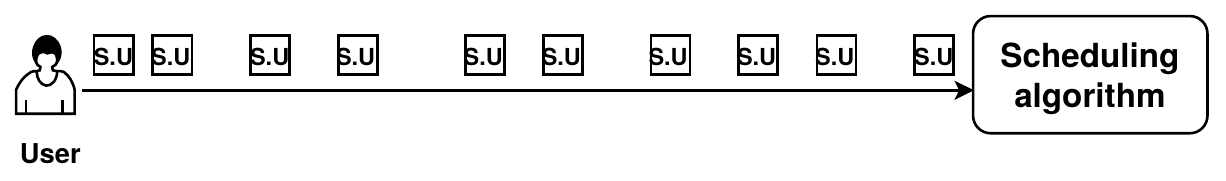}
		\label{fig:tax:app:mode:transactional}
	}\\
	\caption{The top illustrates a user submitting 10 scheduling units in a batch mode, while the bottom  shows submission of 10 scheduling units in a transactional mode.}
	\label{fig:tax:app:mode}
\end{figure*}

Further, when multiple units are submitted for scheduling, all units may arrive at once as a \emph{batch} or may arrive over time, in a \emph{transactional} model, as shown in Fig.~\ref{fig:tax:app:mode}. This \emph{interval of submission} decides the information available to the scheduling algorithm. E.g., in a batch mode, the resource needs and QoS for all units can be used to decide a ``globally optimal'' schedule. But if the units arrive continuously, the scheduler may take individual decisions which can affect the effectiveness of the schedule of future units.

Typically, a batch has a shared QoS requirement defined on it, rather than individual QoS for each unit within.  A batch may also be called a \emph{bag} if there is no specific order in which the units need to be processed. \emph{Ensembles} are a special type of homogeneous batch where all DAGs have a similar or the same structure, but different parameters (e.g., a parameter sweep), number of tasks or task sizes~\cite{Callaghan:2011:MHS:2020806.2020814,5217932}. 
A \textit{bag of tasks (BoT)} consists of a homogeneous batch of tasks which can be executed in any order. BoTs can achieve a high degree of task parallelism and efficiency across VMs and devices since they do not have inter-dependencies~\cite{abramson:ipdps:2000,bots-on-ondemand-vm-2,varshney:tpds:2018}. 

A transactional mode is common when multiple users share the same application deployed on a resource (e.g., FaaS), or a stream of micro-batches arrive from an input source for processing by the application. 
Since requests arrive continuously, future workloads are hard to predict and resources needs to be dynamically allocated and deallocated
~\cite{zhou:tcc:2014,mao:sc:2011,maheswaran:hcw:1999,ghosh2018adaptive}.

\begin{figure}
	\centering
	\includegraphics[width=0.9\columnwidth]{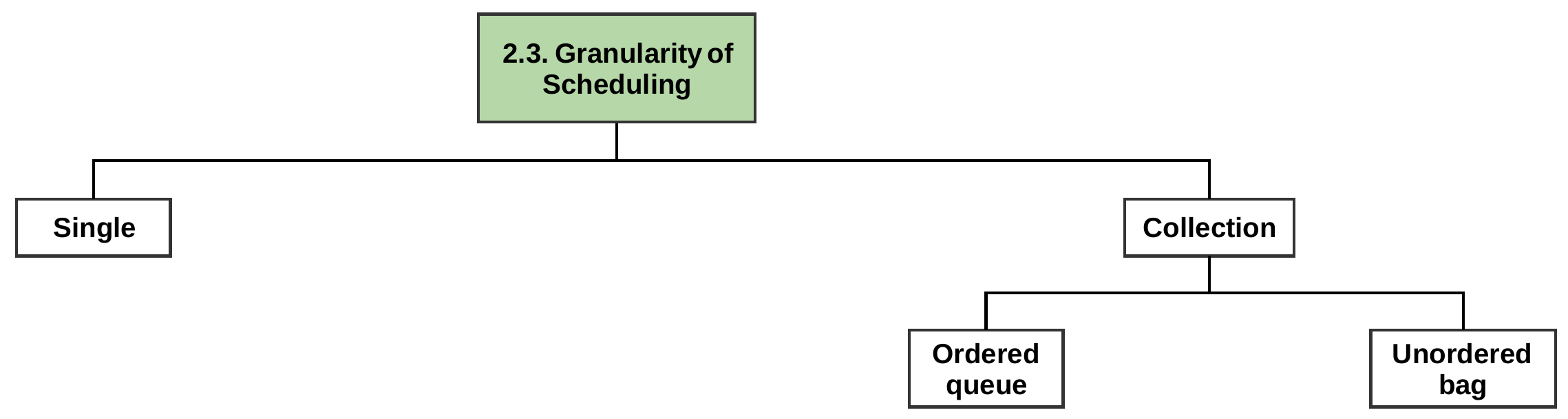}
	\caption{Taxonomy of granularity of scheduling considered by the scheduling algorithm}
	\label{fig:tax:app:gran}

\end{figure}

\subsubsection{Granularity of Scheduling}
\ysnoted{CSUR Reviewer: \emph{``In page 13, granularity of scheduling (" how the algorithm processes the scheduling units after the user submits them") is proposed as a sub-division of application model. Again, it has nothing to do wit the application model and it seems belongs to the scheduling methods.''} This is a fair point. Should this move to Sec 3.4.x?}
The taxonomy for granularity of scheduling is shown in Fig.~\ref{fig:tax:app:gran}, and it identifies the various ways in which an algorithm processes the scheduling units after the user submits them. \emph{Single} unit of scheduling means that the schedule is generated for one unit without considering other units which may be present, while in case of a \emph{collection} granularity, a schedule is generated for a collection of units as a whole, considering the impact of all units within the collection.
This granularity of scheduling is related to the mode of submission. The natural choice is to schedule transaction mode of submissions as single units as they arrive, and batch mode of submissions as collections~\cite{mao:sc:2011,oprescu:cloudcom:2010,bots-on-ondemand-vm-2}.

That said, it is possible 
for a scheduler to buffer units in a transactional workload for a certain period of time and generate a schedule for the collection, thereby increasing the resources efficiency, albeit at a higher latency~\cite{zhou:tcc:2014,xu:ispa:2009}. Similarly, if units within a batch do not have any collective QoS specified on them, the scheduler can ``flatten'' these units and consider then individually for scheduling. 
in~\cite{xu:ispa:2009}, workflows are submitted by the users transactionally, and the ready tasks from the workflows are stored into an ordered queue where the tasks are sorted according to some rules. 
In ~\cite{skarlat2017towards}, IoT applications arrive at any time but the scheduling is done periodically, for applications which are closer to their deadlines. 

\subsubsection{Scheduling unit characteristics}
\label{sec:tax:app:char}

\begin{figure}[t]
	\centering
	\includegraphics[width=0.9\columnwidth]{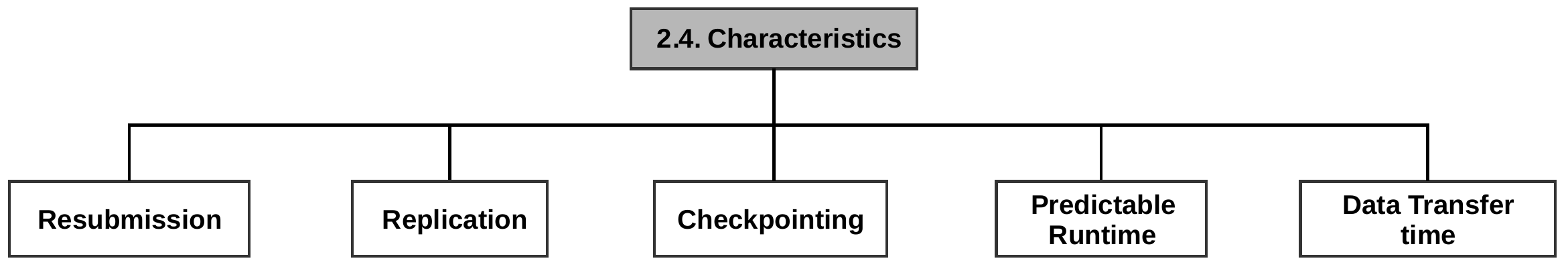}
	\caption{Taxonomy of characteristics of scheduling unit}
	\label{fig:tax:app:char}

\end{figure}

We identify other characteristics of the scheduling unit that can impact the scheduling strategies, 
particularly under dynamic or failure conditions (Fig.~\ref{fig:tax:app:char}).

\ssec{Resubmission} This allows a task (either individually, or as part of a DAG) that has failed to be re-executed completely from the start, without any side-effects. This statelessness or idempotence property is minimally required to ensure fault-tolerance of tasks on unreliable resources~\cite{simmhan:escience:2009}.

\ssec{Replication} This feature allows multiple copies of a task to be run simultaneously on different resources, without any side-effect. This can 
enhance the guarantees for timely completion of a task even if one of the copies fail due to resource loss~\cite{poola:acm:2016,subramanya2015spoton}. This is particularly valuable for pre-emptible cloud resources, or transient or unreliable edge or fog.  It can also be used to opportunistically replicate a task on spare (free) resources so that the first to complete wins, and can address resource under-performance~\cite{calheiros:tpds:2014}.

\ssec{Checkpointing} This allows the scheduler to save the state for a partially executed task, and resume it from the latest checkpoint to meet deadline constraints. 
This can also useful when switching resources to improve the cost or time for execution.
Checkpointing may require migration of the state to a persistent storage (e.g., a cloud storage service) or a different reliable resource before resumption, since the original device or VM may be transient. 

	Checkpointing is leveraged when scheduling on pre-emptible VMs~\cite{yi2010reducing,jung:2014,chu:ipdps:2014,voorsluys:aina:2012}. 
On transient edge devices, CloneCloud uses trigger points to snapshot and migrate local state from the edge to its clone on the cloud to resume execution~\cite{chun2011cloneCloud}. 
\cite{bittencourt:pgcic:2015} migrates the user's data from one Cloudlet to another based on the user mobility, in order to minimize latency~\cite{varshney2017icfec}.

	The \emph{frequency of checkpointing} is important. It balances the progress lost after the last checkpoint when a resource fails 
with the time and cost overhead for taking a checkpoint. 
       Authors have used hourly or user-defined periods~\cite{voorsluys:aina:2012,poola:iccs:2014}, the current and expected spot prices~\cite{chu:ipdps:2014}, the slack time allowed~\cite{subramanya2015spoton}, the mobility and connectivity loss~\cite{shi2012serendipity}, and the job's progress to decide this.
Failure handling can be done at \textit{task level, DAG level} and \textit{bag/ensemble level}. However not all techniques can be used at all levels. E.g., while a workflow may be resubmitted upon failure, it may not be possible to resubmit an entire bag or ensemble. Also, while some critical tasks in a DAG may be replicated, it may be impractical to replicate the entire workflow.

\ssec{Predictable Runtime} The execution time of a task or DAG is deterministic if there is no uncertainty in its expected runtime on different resource sizes. In real world scenarios, the actual runtime might vary due to performance variations and acquisition lag on edge, fog and cloud resources (\S~\ref{sec:tax:sys:char}). The execution times may also change based on specific input parameters. These uncertainties affect the performance of scheduling algorithms~\cite{mao:sc:2011}. 

While many scheduling algorithms rely on the accurate runtime for scheduling units being available~\cite{saeid:fgcs:2013}
, others use initial estimates for the workflow tasks but later use real-time monitoring to revise the execution time and replan the schedule dynamically~\cite{malawski:sc:2012,oprescu:cloudcom:2010,kim:sp:2011,chen:2015}. 

\ssec{Data Transfer time} Workflows and tasks may have large input and output data parameters that need to be transferred between tasks, within a DAG, or between a persistent storage location and the input/output interface to a single task or a DAG. 

These transfers can consume time and cost, both when moved between tasks on different resources in one layer, or between resources on different layers, using local or wide area networks. 
	The application may specify these transfer sizes as part of its definition,  
        and scheduling algorithms may consider them if provided. 
        Data movement also requires storage and network to be available~\cite{varshney2017icfec}. 

In \cite{calheiros:tpds:2014}, VMs are prematurely started 
to allow time to transfer in the data for a task scheduled on it. 
Others include the data transfer time 
to calculate the 
length of the critical path in a DAG~\cite{saeid:fgcs:2013,poola:acm:2016}. 
Some applications that pin the tasks on specific layers, such as the data pre-processing on the edge and analytics on the fog and/or cloud 
may force additional network latencies~\cite{yannuzzi:camad:2014,bonomi2012Fog}.

\begin{figure}[t]
	\centering
	\includegraphics[width=0.9\columnwidth]{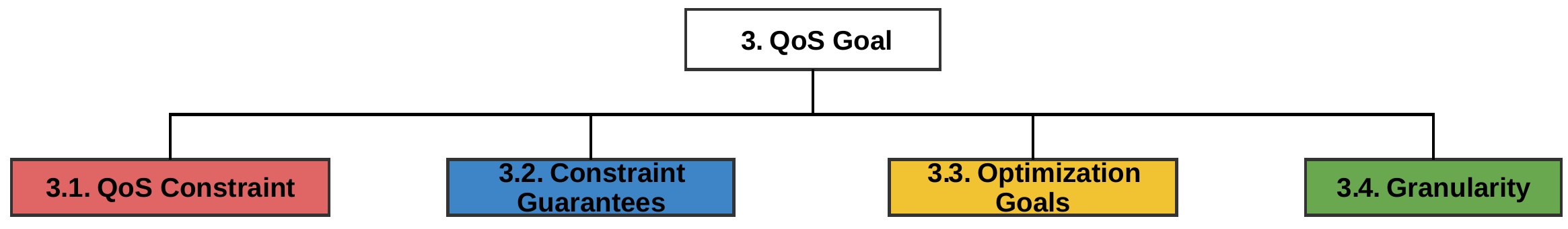}
	\caption{Taxonomy of Quality of Service (QoS) Goal of scheduling algorithm}
	\label{fig:tax:qos}

\end{figure}
\subsection{Quality of Service (QoS) Goals}
\label{sec:tax:qos}

The goal of the scheduling algorithm is to determine a schedule that meets specific Quality of Service (QoS) requirements for the given 

scheduling unit. The QoS is characterized by the type of \textit{constraint} that is imposed, the \emph{guarantees} necessary in achieving the constraints, the \textit{optimization goal} to evaluate the performance, 
and the scheduling \textit{granularity} at which these requirements have to be met, as shown in the taxonomy tree in Fig.~\ref{fig:tax:qos}. 
Besides these, Quality of Experience (QoE) has been proposed as an alternative user-centric metric to QoS~\cite{aazam-2016}. It considers the user requirements and perceptions for a service in a particular context, and calculated using \emph{prediction based} or \emph{feedback-based} approaches. As such, this is a recent evolution, and not considered in our review.

\subsubsection{QoS Constraint}
\begin{figure}[t]
	\centering
	\includegraphics[width=0.75\columnwidth]{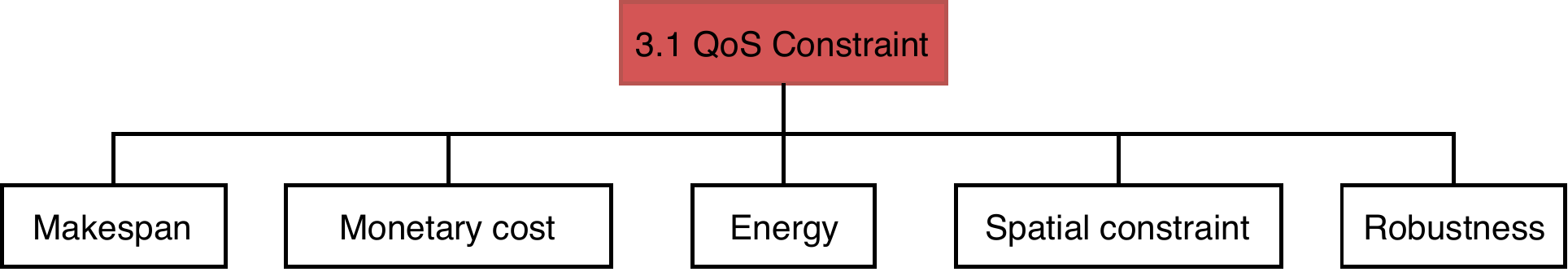}
	\caption{Taxonomy of QoS Constraint specified by user}
	\label{fig:tax:qos:constraint}

\end{figure}

Constraints indicate that the generated schedule must meet the constraint specification, 
while an optimization goal determines the performance of the schedule, provided the constraints are met. 
\emph{time}, \emph{cost}, \emph{energy} and \emph{robustness} are common metrics that are used as constraints and as optimization goals, as seen in Fig.~\ref{fig:tax:qos:constraint}.

\ssec{Time} Temporal constraints are specified on the \emph{makespan} of the scheduling unit, which is the time between it being submitted and it completing execution. Makespan includes any queue waiting time and transfer time, besides the actual task execution time. Users can require that the scheduling unit should complete its execution within a given \emph{deadline} from the submission, based on its importance. 
Related to makespan is the concept of \emph{throughput}, where the number of scheduling units executed per second is the metric. This is relevant for transactional mode of submission where the current or a target rate of requests must be supported. 

\ssec{Cost} For pay-as-you-go resources, the monetary cost is a key factor, and users may specify a \emph{budget constraint} to bound the cost for running the scheduling unit. 
Schedulers may migrate applications to reliable but costlier resources when a deadline is imminent, but end up overpaying. 
So, rather than specify either or both of these constraints independently, users may include a \emph{utility function} that combines both time and cost into a single dimension. The QoS specification can require that this given utility function fall within a certain bound~\cite{Garg:2009:SPA:1862659.1862679,feng2003,saeid:fgcs:2013}.

\ssec{Energy} The overall power consumption by the execution may be specified as a constraint. This may be important for edge devices whose cost may be free but have a limited battery life that should not be exhausted. The energy use may be due to both compute and communication. Edge or fog resources 
on the field may also run on renewables like solar, which have recharge cycles as well that factor into the energy constraint when scheduling~\cite{ghosh2018adaptive}. 

\ssec{Spatial} Geo-spatial constraints may be imposed on applications must be processed on a device close to where the data is generated, e.g., on the same device, the same private network, or some geo-fenced region, to ensure privacy and comply with security policies~\cite{yannuzzi:camad:2014}. 
While the physical security and network access are concerns on edge and fog, the geographical location and legal jurisdiction are factors on the cloud~\cite{brogi2017qos,varshney2017icfec}. These are complementary to resource locality required due to performance. 
Some applications may also pin specific tasks to specific resources, for physical access to on-board sensors or for access to a specific user or device context~\cite{bittencourt:pgcic:2015,Loke2015TheIO,yi:mobidata:2015}.

\ssec{Robustness} Mission critical applications may require 
guaranteed completion. This may pose a higher burden than just completing within a deadline, and may require that failures not happen at all, rather than just be able to recover from failures within the deadline. Examples include 

power grid management or traffic signaling
~\cite{stojmenovic:fedcsis:2014,yannuzzi:camad:2014}.

\subsubsection{Constraint Guarantee}
\begin{figure}[t]
	\centering
	\includegraphics[width=0.75\columnwidth]{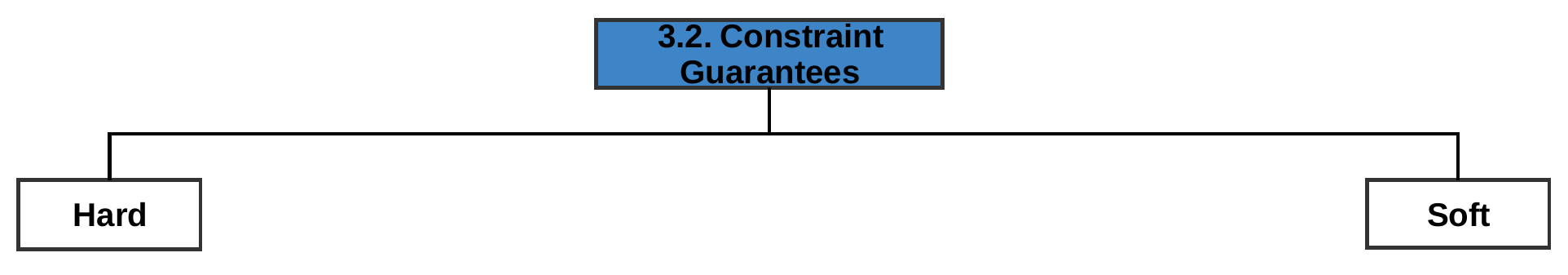}
	\caption{Taxonomy of Constraint Guarantees specified by user}
	\label{fig:tax:qos:guarantee}

\end{figure}

The constraint specified by the user can be \emph{hard} or \emph{soft} (Fig.~\ref{fig:tax:qos:guarantee}). A hard constraint is inviolable, and its failure is catastrophic for the end-user. 
\cite{jin2012network} refers to applications with hard time constraints as \emph{inelastic applications}, and they require real-time processing, typically to meet safety of humans, such as in autonomous vehicles~\cite{Bonomi2014}. Soft requirements, on the other hand, need not strictly be achieved and a best effort is sufficient~\cite{mao:sc:2011}. In such cases, penalty functions may be used when the constraints are not satisfied. 
Such \emph{elastic applications} (different from cloud elasticity) have flexibility in latencies and 
can perform batch processing, e.g., for analyzing surveys from drones~\cite{Loke2015TheIO,varshney2017icfec}.
In scheduling literature, the auto-scaling mechanism in \cite{mao:sc:2011} considers soft deadlines for the workflows while RTBA~\cite{chu:ipdps:2014} enforces hard deadlines. LOSS and GAIN use hard budget constraint to find a schedule that gives the shortest makespan~\cite{sakellariou:coregrid:2007}.

\subsubsection{Optimization Goal}

\begin{figure}[t]
	\centering
	\includegraphics[width=0.85\columnwidth]{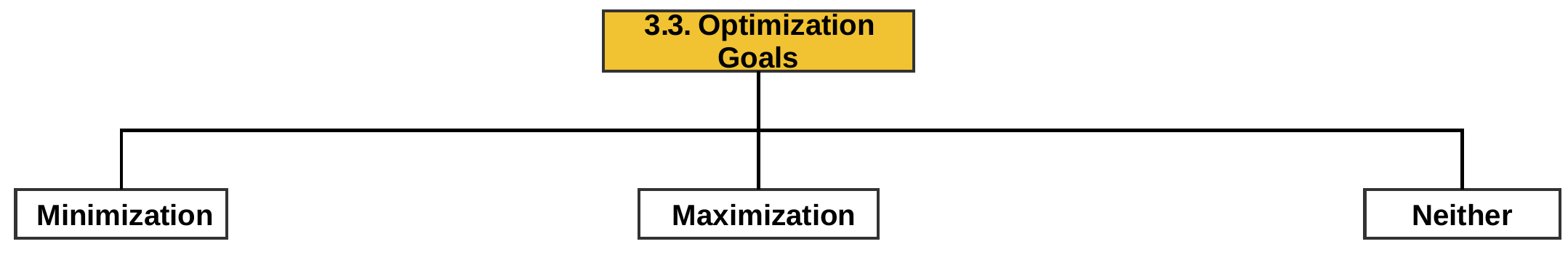}
	\caption{Taxonomy of Optimization Goals of scheduling algorithms}
	\label{fig:tax:qos:goal}

\end{figure}

The QoS optimization goal attempts to minimize or maximize an \emph{objective function} for the application's schedule, as shown in Fig.~\ref{fig:tax:qos:goal}. The metric for the objective function is similar to the ones for the QoS constraint, such as the makespan, throughput, monetary cost, energy, utilization, faults, or a functional combination of these.
\ysnoted{CSUR Reviewer: \emph{``In page 18, I think dividing Optimization Goals to maximization and minimization is not interesting, because every optimization problem in the world belongs to one of these categories. Maybe it is better to divide Optimization Goals to the criteria such as makespan, throughput and ...''} One can argue both ways...Maybe we need both dimensions: makespan, throughput, monetary cost, energy, faults, utilization}

\ssec{Minimization} Typically, the makespan, the cost or energy consumption, or the number of task failures are used as the objective function when minimizing it. A deadline constrained application may attempt to minimize the cost, with a number of such scheduling algorithms available~\cite{saeid:fgcs:2013,rodriguez:tcc:2014,mao:sc:2011,zhou:tcc:2014,poola:acm:2016,chu:ipdps:2014}. Similarly, under a budget constraint, the makepsan may be the minimization goal~\cite{sakellariou:coregrid:2007,bots-on-ondemand-vm-2}. Others may lack a constraint, and only aim to minimize, say, the makespan, the energy consumption, or both time and cost~\cite{topcuoglu:tpds:2002,hong:mcc:2013,shi2012serendipity,chun2011cloneCloud,deng2016optimal,ghosh2018adaptive,wu:2013,fard:tpds:2013}.

\ssec{Maximization} Common objective metrics used when maximizing include the application throughput and the utilization of resources.~\cite{skarlatprovisioning, skarlat2017towards} maximize the utilization of cheap fog resources by trying to place more applications on it instead of the cloud to reduce the monetary cost.
	 \cite{wang:ijcsns:2006} schedules BoTs to maximize the survivability given a deadline constraint, with task priorities 
used as a proxy for user-specific maximization goals. 

\cite{malawski:sc:2012} maximizes the number of high-priority workflows from an ensemble that complete, given a fixed budget and deadline constraint.

\ssec{Neither} It is also possible that no optimization goal is specified. In \cite{calheiros:tpds:2014}, the user specifies a preferred deadline and a variable budget, and the goal is to meet the soft deadlines of workflows. Kushwaha, et al.~\cite{kushwaha:ccem:2014} analyze the trade-offs between cost savings over fixed price VMs and resilience when tasks are run on spot VMs. 
In~\cite{brogi2017qos}, users specify the hardware, software, bandwidth and latency requirements for tasks, and aim to enumerate all valid deployments on the fog and cloud.

\subsubsection{Granularity}
The QoS optimization goals and constraints can be specified at various granularities, with the default being at the granularity of the single scheduling unit or the batch, depending on the mode of submission. 

\begin{figure}[t]
	\centering
	\includegraphics[width=0.85\columnwidth]{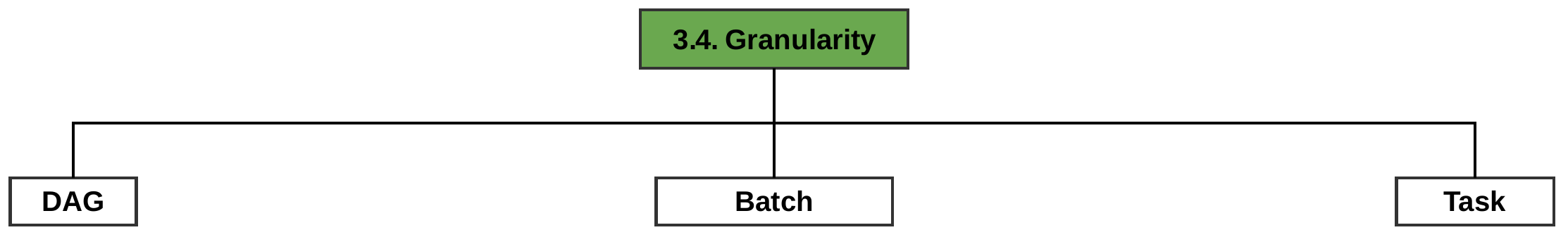}
	\caption{Taxonomy of Granularity of QoS Goals}
	\label{fig:tax:qos:gran}

\end{figure}

However, other variations in specifying the granularity of constraints and optimizations exist as well, as seen in 
Fig.~\ref{fig:tax:qos:gran}.

\ssec{DAG} Placing the QoS on the DAG means that it has to be met for the DAG as a whole without regard to the QoS for individual tasks. In such cases, the constraints and the optimization goals are specified at the same granularity~\cite{rodriguez:tcc:2014,poola:acm:2016,saeid:fgcs:2013}. However, these may be different as well. E.g., the makespan constraint may be specified at the DAG level, but the goal of minimizing cost specified for a \emph{batch} of DAGs~\cite{zhou:tcc:2014,mao:sc:2011,skarlat2017towards}.

	When constraints are specified for a DAG, the tasks lying on its critical path are essential to manage the makespan of the workflow. Scheduling algorithms may estimate and assign sub-deadlines for each task to decide their mapping to a suitable resource, with tasks in the critical path having the least flexibility. E.g., the IC-PCPD2 algorithm distributes the sub-deadline of a path in a DAG to each task in the path, in proportion to its minimum execution time~\cite{saeid:fgcs:2013} while~\cite{liu:hpca:2010} uses the average execution time. 
	
        Sub-deadlines can be used to 
select the best resource that can execute the task within its sub-deadline. In Serendipity~\cite{shi2012serendipity}, sub-deadlines on each task are assigned to minimize the overall time and the energy consumed for executing the DAGs.

\ssec{Batch} QoS can also be specified on a batch of scheduling units, be they a Bag of Tasks or an ensemble of workflows. The constraints and/or goals have to be met for the entire collection, without regard to individual units. 
E.g., 
Varshney and Simmhan~\cite{varshney:tpds:2018} define a deadline constraint and a cost minimization goal to schedule a Bag of Tasks executing on pre-emptible and on-demand VMs.
It is possible that the scheduler may not compute the sub-constraints for the individual units of the batch~\cite{malawski:sc:2012}. 
In~\cite{deng2016optimal}, a delay constraint is specified for the batch of task and the goal is to minimize the overall power consumption on fog resources. 
	
        ~\cite{skarlatprovisioning} periodically schedule a batch of tasks with 
the goals of minimizing the makespan for the batch and maximizing the resource utilization.

\ssec{Task} The user may also specify goals or constraints for individual tasks, be they tasks of a DAG, tasks within a bag or single tasks~\cite{chu:ipdps:2014,kushwaha:ccem:2014,voorsluys:aina:2012}. It is also possible for users to indicate the constraints and optimization goals individually for different stages of the DAG. In such cases, different algorithms can be used to schedule the tasks of different stages according to the constraints and goal at each stage~\cite{diaz:tcc:2015}

\subsection{Scheduling Algorithms}
\label{sec:tax:algo} 
Given the edge, fog and cloud resource environment, and the application model and QoS requirements, the scheduling algorithm is designed to schedule the application onto the resources to meet the QoS goal and meet the constraints. Solving this scheduling problem is computationally complex for non-trivial problem sizes. As a result, there is a large body of literature on techniques and algorithms to solve this problem, often to get an approximate rather than an optimal solution. Existing surveys classify scheduling algorithms based on various techniques that they employ, and these determine the quality of the schedule that is generated, and the time taken to generate the same~\cite{wu:2015,yu:buyya:survey:grid,huang:jsw:2013,zhan:csur:2015,liu:ccgrid:2014}. 

\begin{figure}[t]
	\centering
	\includegraphics[width=0.99\columnwidth]{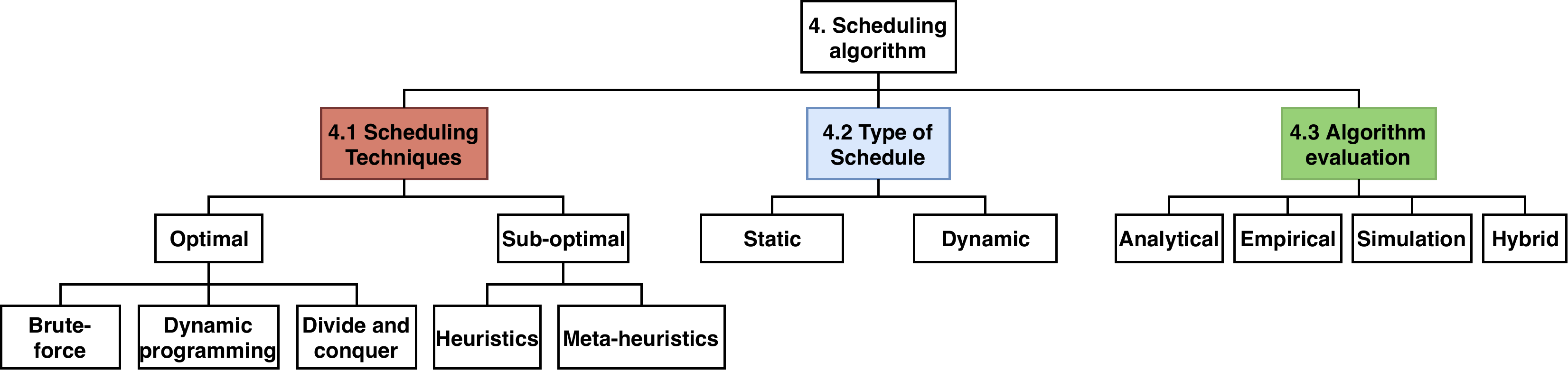}
	\caption{Taxonomy of the techniques used and types of schedules generated by scheduling algorithms}
	\label{fig:tax:algo}

\end{figure}

As such, there are no intrinsic reasons why these prior classifications, as illustrated in Fig.~\ref{fig:tax:algo}, are not equally applicable to application scheduling on edge, fog and clouds. However, having multiple resource layers across a wide area network also brings in the opportunity for delegating the scheduling hierarchically (e.g., cloud delegating a subset of the scheduling unit to a fog for scheduling on itself and its neighboring edges)~\cite{yannuzzi:camad:2014,stojmenovic:fedcsis:2014,yi:mobidata:2015,hong:mcc:2013}, or to make federated decisions~\cite{vaquero2014finding}, rather than just a centralized decision. We examine these and, to be holistic, also summarize outcomes from prior taxonomies in this section; we refer readers to these external sources for a detailed review of the algorithmic strategies.

\subsubsection{Scheduling Techniques}\label{sec:tax:algo:tech}

Finding the \emph{optimal schedule} is a combinatorial optimization problem and a variation of the classic ``job shop scheduling'' problem, the solution to which is NP Hard~\cite{yamada:1997:gaes}. \emph{Brute force algorithms} try all possible mappings of the scheduling units to the compute resources to arrive at a globally optimal solution while meeting the constraints. E.g., a simple case of optimally placing $M$ tasks to $R$ resources has a brute force computational complexity of $\mathcal{O}(R^M)$. This is intractable for practical applications with $10-100's$ of tasks and resources~\cite{Convolbo2016,ghosh-2018}.

In some cases, one can stop the brute force search once a ``good enough'' solution is found and thus bound the cost of the algorithm, while not getting an optimal solution. 

There are also approaches that use \emph{dynamic programming (DP)}, wherein the original scheduling problem is decomposed into a set of \emph{overlapping} sub-problems, each of which can be solved optimally in tractable time. Then the solutions to the sub-problems are directly reused using ``memo\-iza\-tion'' when exploring the search space. E.g., RTBA uses DP to construct a strategy table for a task of a given size and deadline constraint, which is used to schedule all tasks that fall within this size and deadline~\cite{chu:ipdps:2014}. This has a time complexity of $\mathcal{O}(C^2T_{dl})$ where $T_{dl}$ is the deadline and $C$ is the compute time, 
and this is faster than the brute force approach which has a complexity of $\mathcal{O}(|A|^{T_{dl}})$, where $A$ is the set of possible scheduling actions. 
As we see, DP gives optimal solutions much faster than brute force. But it is still unlikely to be fast enough for practical use in online scheduling.

\emph{Divide and conquer algorithms} partition the scheduling problem into smaller \emph{non-overlapping} sub-problems, and solve these sub-problems recursively. The solutions to the sub-problems (tasks) are combined into a solution to the original problem.

In practice, such optimal solutions are used for small problem that then contribute to an overall approximate solution.~\cite{deng2016optimal}
decomposed the overall problem of distributing a workload among fog and cloud resources into sub-problems using an approximate, and solve these sub-problems individually using different optimization techniques.

Alternatively, the optimal solution, while impractical, offers a theoretical baseline against which to empirically compare their proposed approximate solution~\cite{Convolbo2016,chu:ipdps:2014}.

Consequently, scheduling problems are often solved using \emph{sub-optimal} algorithms that resort to heuristics or meta-heuristics that, in practice, come close to optimal solutions but within a reasonable time.  \emph{Greedy algorithms} select the most promising option from the solution space at any given stage. It makes a series of locally optimal choices with the expectation of finding the global optimum. 
Greedy DAG scheduling heuristics such as HEFT schedule DAGs over Grid resources~\cite{topcuoglu:tpds:2002}, and have been reused for scheduling on heterogeneous cloud VMs to give short makespan~\cite{zheng:jgc:2013,lin:cloud:2011}. HEFT has a complexity of $\mathcal{O}(E.R)$ where $E$ is the number of edges in the DAG and $R$ is the number of resources. 
These have also been extended to include cost/time budgets/goals using functions like GAIN and LOSS
~\cite{sakellariou:coregrid:2007,mao:sc:2011,diaz:tcc:2015}. 

There are greedy algorithms for scheduling BoTs on cloud resources~\cite{bots-on-ondemand-vm-1} and DAG scheduling on edge devices~\cite{shi2012serendipity,ghosh2018adaptive,brogi2017qos}. They use heuristics such as predicting the proximity between mobile edge devices for deadline planning, 
prioritizing tasks with the most resource constraints,  
and incrementally co-locating tasks on the same resource to avoid network latency. 
More sophisticated heuristics have also been proposed, such as the ToF planner and AutoBoT that uses a greedy approach to minimize the monetary cost for executing a bag of workflows or tasks on clouds~\cite{zhou:tcc:2014, varshney:tpds:2018}.

Similarly, \emph{backtracking} algorithms follow one of many possible alternatives, and backtrack if it does not look promising~\cite{chen:2015,brogi2017qos}.

Workflow scheduling problems are often reduced to \emph{Integer Linear Programming (ILP)} and its mixed variant~\cite{chun2011cloneCloud}. E.g., CloneCloud~\cite{chun2011cloneCloud} uses ILP to 
find partition points between edge and cloud in the application which minimize the overall execution time or energy consumption. 

While standard techniques exist to provide optimal solutions to ILP problems, they can only be used for small size problems due to their prohibitive computational cost. So, heuristics are used to solve these ILP problems as well~\cite{diaz:tcc:2015,jung:iscc:2011,malawski:sp:2015}.

\emph{Meta-heuristics} are a class of high level guidelines that can be used to define heuristics to solve a wide class of optimization problems~\cite{blum:2003:csur}. Specific scheduling problems are refactored to fit the higher-order problem after which the heuristic guidelines can be applied to explore the solution space. Meta-heuristics have been categorized into trajectory-based and population-based methods, with \emph{simulated annealing} falling in the former category, and \emph{genetic algorithm (GA)}, \emph{particle swarm optimization (PSO)} and \emph{ant colony optimization (ACO)} featuring in the latter method. 

E.g., in \cite{rodriguez:tcc:2014}, a deadline-constrained workflow scheduling on clouds is mapped to a PSO problem. A particle's coordinates encode the mapping between the task and a resource, and the dimension of the particle matches the number of tasks. The fitness function for the PSO, which has to be minimized, is the total execution cost, and the schedule is generated by solving this PSO problem with a complexity of $\mathcal{O}(P.M^2.R)$ per iteration, where \textit{P} is the number of particles, \textit{M} is the number of tasks and \textit{R} is the number of resources. In~\cite{ghosh-2018}, a DAG scheduling problem across edge and cloud has been reduced to GA formulation. 
Each chromosome represents a mapping function from a task to an available edge or cloud resource. The chromosome which gives the minimum makespan, among all the solutions that do not violate compute and energy constraints, is selected from across generations.

Scheduling heuristics are also tuned for operating across resource abstraction layers, such as tasks across mobile edge and cloud layers~\cite{chun2011cloneCloud}, or DAGs across fog and cloud~\cite{skarlat2017towards}. Some also use locality of tasks to layers. Bittencourt, et al.~\cite{bittencourt2017mobility} schedule DAGs that are submitted to a Cloudlet, and scheduler can execute it on the ``local'' or a remote fog, and/or cloud resources. 

Likewise, Ghosh and Simmhan~\cite{ghosh2018adaptive} schedule transactional DAGs on edge and cloud resources, but pin the source tasks to the edge and the sink tasks to the cloud.

\subsubsection{Type of Schedule} \label{sec:tax:algo:sched} 
The scheduling algorithm can be designed as a \emph{static} (offline) algorithm that is run once when the scheduling unit arrives, or a \emph{dynamic} (online) algorithm that actively decides the schedule based on the current conditions through the lifetime of the unit.

\ssec{Static schedule} In a static approach, the mapping from the scheduling unit to resource(s) is generated once before the unit starts, based on the information about other units and the resources available to the scheduling algorithm \emph{a priori}. This allocation is retained for the lifetime of the task or DAG, and does not respond to changes in the resources or tasks at runtime. It assumes that the prior knowledge is perfect. It also does not make use of checkpointing and resubmission.

Abrishami, et al.~\cite{saeid:fgcs:2013} generate an offline schedule for a deadline constrained DAG. It assigns a partial deadline to tasks in the critical path of the DAG, 
along with their \emph{Earliest Start Time (EST)}, 
\emph{Earliest Finish Time (EFT)}, 
and \emph{Latest Finish Time (LFT)}
~\cite{topcuoglu:tpds:2002}. 
The algorithm then recursively assigns all the tasks on the path to a single VM which can finish each task before its LFT, with the minimum price.

SPSS also uses static scheduling to schedule an ensemble of workflows~\cite{malawski:sc:2012}. They use \emph{admission control} to prevent scheduling of workflows which cannot complete within the deadline and budget constraints.

While static algorithms do not consider runtime dynamism, \cite{calheiros:tpds:2014} mitigates the effects of variations in performance using task replication. 
It uses the budget surplus and the idle slots in the allocated resources, after performing a static schedule, for task replication. This increases the chances of the deadline being met.

Static scheduling is also possible for a transactional mode of submission. Many cloud providers including Amazon, Google and Azure offer users \emph{rule based auto-scaling} for transactional tasks. Here, incoming tasks are routed, typically, in a round-robin manner to a pool of active VMs based on user rules that decide when and how to increase or decrease the number of VMs in this pool based on their utilization. 
These can efficiently handle the dynamic task-based transactional workloads if the user is able to select the right threshold value~\cite{mao:sc:2011,nathan:icac:2013}. Similarly, sensor and event based applications have a dynamic workload pattern that schedulers adapt to on edge and fog resources~\cite{hong:mcc:2013,ghosh2018adaptive}. 

However, some auto-scaling mechanisms use future workload prediction by monitoring current resource utilization to perform proactive auto-scaling~\cite{morais:ccgrid:2013}.

Static schedules are useful when the workload is known in advance and the performance of resources is deterministic. Real-time applications, those with a closed control loop, and those sensitive to mobility require careful design. These may prefer a static schedule to ensure determinism, preclude the use of mobile resources, and/or retain the decision logic in a single resource~\cite{sadeghi:hipc:2016}. However, static planning, exclusively, cannot respond to faults and to dynamism in compute and network performance, acquisition time, and spot prices on edge, fog and cloud resources. These can cause sub-optimal QoS performance or, worse, violation of hard constraints.

\ssec{Dynamic schedule} Dynamic algorithms use runtime knowledge about the resource performance and application behavior, besides static knowledge, to make scheduling decisions at the start and during the lifetime of the scheduling unit.

This information helps them adapt their schedule on the fly to avoid QoS constraint violations, and/or to improve the optimization goals. 
Dynamic algorithms can either be \emph{just-in-time}, where the scheduling decision for a task in a workflow is made just once before the task's execution (but after the DAG itself has started executing), or \emph{fully dynamic} where running tasks can be remapped from one resource to another during its execution. 

DPDS maintains a priority queue of ready tasks from a batch of DAGs, ordered by the workflow priority, that are scheduled \emph{just-in-time}~\cite{malawski:sc:2012}.

Whenever a VM is idle, the ready task at the head of the queue is assigned to it.  
Once a task completes, other dependent tasks in the DAG which are ready are added to this queue and the scheduling continues.

A similar strategy is used for scheduling multiple workflows that are submitted in a transactional mode~\cite{xu:ispa:2009}.

The tasks in the queue are mapped just-in-time to the best VM to ensure sub-deadline and sub-budget are not violated.

Other just-in-time heuristics schedule workflow tasks onto both spot and on-demand VMs~\cite{poola:iccs:2014}. At runtime, it decides for each ready task whether to schedule it on a spot VM or an on-demand VM based on its \emph{Latest Time to On-demand (LTO)}, i.e., the slack time before which they must be executed, to avoid exceeding the deadline. 
Tasks that are ready before their LTO are scheduled on spot VMs while ones that arrive after their LTO are mapped to on-demand VMs.

Elsewhere, the progress of each task is continuously monitored and if one gets delayed, it is dynamically rearranged and future tasks rescheduled to prevent an increase in the overall makespan of the application~\cite{chen:2015}.

RTBA uses a \emph{fully dynamic algorithm} which actively manages the life-cycle of individual tasks on spot and on-demand VMs~\cite{chu:ipdps:2014}. It statically constructs a strategy table, as discussed earlier, 
and performs dynamic lookups on this table to decide actions based on the current spot price and task progress. 
The actions can checkpoint and migrate the task, change the bid price, or even pause the task for the bid price to drop.

Similarly, CloneCloud~\cite{chun2011cloneCloud} initially builds a partition database for the application, and a partition configuration is selected from this at runtime based on the current resource availability and network conditions.

The ToF planner combines both \emph{static and dynamic strategies} to schedule transactional workflows~\cite{zhou:tcc:2014}. It accumulates workflows arriving for a certain time period and statically assigns a 
provisional set of VMs 
to their tasks. However, the VM instance itself is only decided (or started) at runtime to enable the reuse of running VMs.

Applications that are scheduled on edge and fog resources need to be particularly responsive to resource dynamism that are more acute on these platforms~\cite{varshney2017icfec,shi2012serendipity}. 
This also requires the runtime capabilities for monitoring to determine when such adaptation is required~\cite{dasterdi:corr:2016}. This may even require \emph{changing the coordination strategy} from, say, centralized to a federated or P2P one.

\subsubsection{Algorithm Evaluation}\label{sec:tax:algo:eval}

Lastly, we consider how the proposed scheduling algorithm is evaluated for its ability to meet the QoS for the application and system models. Others~\cite{coutinho:2015} already identify some of these categories of evaluations, such as experimental, analytical, simulation and combinations of these. These used in our classification for completeness.

Some papers offer both \emph{analytical proofs} and experimental validation to measure the efficiency of their scheduling algorithm~\cite{fard:tpds:2013}. 

Several others consider only an \emph{empirical evaluation} using real world scientific applications like Cybershake, Genome, LIGO, and Montage~\cite{malawski:sc:2012,saeid:fgcs:2013,poola:iccs:2014,zhou:tcc:2014}, or domain-specific workloads~\cite{riotbench}. For scheduling across edge, fog and cloud, mobile applications such as virus scanning, image search and video analytics have been used for empirical evaluation~\cite{shi2012serendipity,chun2011cloneCloud,khochare:iscocw:2017}. But the predominant choice of validation in literature is limited to \emph{simulations} based on synthetically generated BoTs and DAGs~\cite{bots-on-ondemand-vm-1,lin:cloud:2011,ghosh2018adaptive}. 

During simulation, various probability distributions can be used to generate synthetic BoTs and to model the arrival rates of tasks in case of transactional mode~\cite{bots-on-ondemand-vm-1,maheswaran:hcw:1999,ghosh2018adaptive,shi2012serendipity}. Random DAG generators are also used to generate DAGs with parameters and distributions to control the number of tasks, number of dependency edges, range of task length, depth and width of DAG, outdegree, and communication to computation ratio~\cite{chen:2015,zeng:aina:2012,wu:2013}.

Alternatively, publicly available cloud workloads, such as the Google Cluster Workload also offer the choice of realistic DAG and BoT simulations~\cite{reiss:Cloud:2012,varshney:tpds:2018}. The CloudSim~\cite{calheiros:spe:2011} and iFogSim~\cite{fogsim} simulation frameworks can model application execution on cloud infrastructure, and edge and fog infrastructure respectively, and are often used to evaluate scheduling algorithms~\cite{poola:iccs:2014,malawski:sc:2012,skarlatprovisioning,skarlat2017towards,bittencourt2017mobility}. 

Many algorithms consider system models that are \emph{realistic}, and 
similar to those provided by major cloud providers. Amazon AWS is a popular choice here, with many considering different EC2 on-demand VM types, and their prices and configurations in their simulation~\cite{mao:sc:2011,calheiros:tpds:2014,poola:acm:2016}. 
Some~\cite{ghosh2018adaptive,shi2012serendipity} also use real-world benchmarks and distributions of network and compute performance of edge and cloud resources.

Algorithms which use pre-emptible VMs simulate the fluctuation in spot prices using historic spot price data provided by Amazon~\cite{poola:iccs:2014,varshney:tpds:2018}, and to train their price bidding models and expected lifetime estimation models~\cite{chohan:usenix:2010,chu:ipdps:2014}. The acquisition lag for VMs may be based on a distribution of past startup times~\cite{Schad:2010:RMC:1920841.1920902,rodriguez:tcc:2014}, or limited to a constant value~\cite{malawski:sc:2012,mao:sc:2011}. Similarly, the mobility of edge and fog devices can be captured using real-world mobility traces or generated synthetically using mobility models~\cite{shi2012serendipity}.~\cite{hong:mcc:2013} simulate random vehicle motion for a vehicle-to-vehicle streaming application, where each vehicle randomly picks another vehicle to start streaming a video to it.

\begin{sidewaystable*}
\scriptsize
\centering

\caption{{\small Classification of Scheduling Algorithms}}

\label{tbl:review:1} 
\begin{tabular}{p{0.7cm}||p{0.7cm}||p{0.95cm}|p{0.6cm}|p{1.5cm}||p{0.65cm}|p{1.1cm}|p{1.9cm}||p{1cm}|p{1cm}|p{0.8cm}||p{1.6cm}|p{1.0cm}|p{2.2cm}}

\hline
\textbf{Paper} 
		& \multicolumn{4}{c||}{\textbf{System Design}~\S~\ref{sec:tax:sys}}  
& \multicolumn{3}{c||}{\textbf{Application Model}~\S~\ref{sec:tax:app}} 
& \multicolumn{3}{c||}{\textbf{Quality of Service}~\S~\ref{sec:tax:qos}} 
& \multicolumn{3}{c}{\textbf{Algorithm}~\S~\ref{sec:tax:algo}} 
\\ \hline
		& Layer$^1$ & Size$^2$ & Price$^3$ & Char.$^4$
		& Struct. & Mode$^5$ & Char.
		& Constr.$^6$ & Goal$^7$ & Gran.$^8$ 
		&  Technique & Schedule & Evaluation$^9$ \\ \hline
\hline
		\cite{saeid:fgcs:2013} & C & HT & OD & BI-60,5~min, Linear PPR
& DAG & S & Data transfer time
& TD/H & MIN(C) & DAG, DAG
& Divide and Conquer & Static & S -- \emph{Synthetic DAGs of CyberShake, Epigenomics, LIGO, Montage, SIPHT}
\\
\hline
		\cite{bittencourt:jisa:2011} & C & HT & OFF, OD & BI-60~min
& DAG & S & Data transfer time
& TD/S & MIN(C) & DAG, DAG
& Greedy & Static & S -- \emph{Synthetic DAGs of AIRSN, Chimera, CSTEM, LIGO, Montage, random}; E -- \emph{Image processing application} \\
\hline
		\cite{calheiros:tpds:2014} & C & HT & OD & BI-60~min, AL, PV
& DAG & S & Resubmit, Replicate, Data transfer time
& TD/S and CB/H & -- & DAG, DAG 
& Heuristic & Static & S -- \emph{Synthetic DAGs of Montage, CyberShake, LIGO, SIPHT} \\
\hline
		\cite{chen:2015} & C &  HT & -- & Limited availability
& DAG & \{M, HO, T\}/S & Non-deterministic runtime, Data transfer time
& TD/S or -- & MAX(O) or MIN(T) & DAG, Batch 
& Backtracking & Dynamic & S -- \emph{Random DAGs} \\
\hline
		\cite{chohan:usenix:2010} & C & HO & OD, SP & BI-60~min, NP \emph{(asymm.)}
& DAG & S & --
& -- & MIN(T) & --, DAG 
& Heuristic & Static & E -- \emph{WordCount, Pi, Sort on EC2} \\
\hline
		\cite{chu:ipdps:2014} & C & OD:HT, SP:HT, OFF:HO & OD, SP, OFF & BI-60~min, NP \emph{(asymm.)}, AL
& Task & S & Checkpoint, Migrate, Data transfer time
& TD/H & MIN(C) & Task, Task 
& Dynamic Programming & Dynamic & S -- \emph{Varying length tasks} \\
\hline
		\cite{fard:tpds:2013} & C & HT & OD & BI \emph{(novel pricing model)}, PV
& DAG & S & Data transfer time
& -- & MIN(C) and MIN(T) & DAG, DAG 
& Greedy & Dynamic & H -- \emph{Synthetic WIEN2k DAG, Random DAGs} \\
\hline

		\multicolumn{14}{l}{$^1$~E: Edge, F: Fog, C: Cloud, M: Spatial mobility} \\
		\multicolumn{14}{l}{$^2$~HO: Homogeneous machine sizes, HT: Heterogeneous sizes} \\
		\multicolumn{14}{l}{$^3$~OD: On-demand fixed price VMs, SP: Pre-emptible spot/variable price VMs, OFF: Off-Cloud resources, FR: Free Edge/Fog resources}\\
		\multicolumn{14}{l}{$^4$~BI: Billing Interval, AL: Acquisition Lag considered, PV: Performance Variations considered, NP: Network Pricing, PPR: different Price to performance ratios considered}\\
		\multicolumn{14}{l}{$^5$~S: Single scheduling unit submitted, M: Multiple units; HO: Homogeneous units, HT: Heterogeneous units; B: Batch submission, T: Transactional submission;}\\
\multicolumn{14}{l}{\quad(S): Single granularity of scheduling, /O: Ordered queue, /U: Unordered bag.}\\
		\multicolumn{14}{l}{$^6$~TD: Time deadline, CB: Cost budget, E: Energy constraint, S: Spatial constraint /H: Hard constraint, /S: Soft constraint}\\
		\multicolumn{14}{l}{$^7$~MIN: Minimize objective function, MAX: Maximize function, (T): Time, (C): Cost, (O): Other.}\\
		\multicolumn{14}{l}{$^8$~Gran.: Granularity of QoS constraint, Granularity of QoS Goal}\\
		\multicolumn{14}{l}{$^{9}$~A: Analytical, E: Empirical, S: Simulation, H: Hybrid} \\
\end{tabular}
\end{sidewaystable*}

\begin{sidewaystable*}
\centering
	\scriptsize

{\small Table 1. Classification of Scheduling Algorithms \emph{(continued)}}\\

\begin{tabular}{p{0.7cm}||p{0.7cm}||p{0.95cm}|p{0.6cm}|p{1.5cm}||p{0.65cm}|p{1.1cm}|p{1.9cm}||p{1cm}|p{1cm}|p{0.8cm}||p{1.6cm}|p{1.0cm}|p{2.2cm}}

			\hline
			\textbf{Paper} 
		& \multicolumn{4}{c||}{\textbf{System Design}~\S~\ref{sec:tax:sys}}  
			& \multicolumn{3}{c||}{\textbf{Application Model}~\S~\ref{sec:tax:app}} 
			& \multicolumn{3}{c||}{\textbf{Quality of Service}~\S~\ref{sec:tax:qos}} 
			& \multicolumn{3}{c}{\textbf{Algorithm}~\S~\ref{sec:tax:algo}} 
			\\ \hline
		& Layer$^1$ & Size$^2$ & Price$^3$ & Char.$^4$
		& Struct. & Mode$^5$ & Char.
		& Constr.$^6$ & Goal$^7$ & Gran.$^8$ 
		&  Technique & Schedule & Evaluation$^9$ \\ \hline
			\hline
		\cite{bots-on-ondemand-vm-1} & C & HT & OD & BI-60~min
& Task & \{M, HO, B\}/O$^\dagger$; \{M, HO, T\}/S$^\ddagger$ & --
& -- & MIN(T) and MIN(C) & --, Batch (overall BoTs) 
& Greedy & Dynamic & S -- \emph{Random BoTs} \\
\hline
		\cite{kim:sp:2011} & C & HT & OFF, OD & BI-60~min, AL, PV, Failures 
& Task & \{M, HO, T\}/O & Resubmit, Non-deterministic runtime
& TD/S or CB/S & MIN(T) or MAX(O) & Batch, Batch
& Greedy & Dynamic & E -- \emph{DAG of Reservoir simulation ensemble, Kalman filter on TeraGrid, EC2} \\ 
\hline
		\cite{kushwaha:ccem:2014} & C & HT & SP & BI-60~min
			& Task & S & --
			& -- & MIN(C) & --, DAG 
			& Heuristic & Static & S -- \emph{Varying length tasks} \\
			\hline
		\cite{lin:cloud:2011} & C & HT & OD & --
			& DAG & S & Data transfer time
			& -- & MIN(T) & --, DAG
			& Greedy & Static & S -- \emph{Random DAGs} \\
			\hline
		\cite{liu:hpca:2010} & C & HT & OD & BI-N/A, Linear PPR, Failures, Limited availability
			& DAG & \{M, HO, B\}/O & Resubmit, Data transfer time
			& TD/S & MIN(C) & Batch, Batch
			& Divide and Conquer & Dynamic & S -- \emph{Synthetic DAGs based on practical workflows such as bank cheque workflow processing} \\
			\hline
		\cite{malawski:sc:2012} & C & HO & OD & BI-60~min, AL
			& DAG & \{M, HO, B\}/O & Non-deterministic runtime
			& TD/H and CB/H & MAX(O) & Batch, Batch 
			& Heuristic (DPDS, WA-DPDS), Divide and Conquer (SPSS) & Dynamic (DPDS, WA-DPDS), Static (SPSS)& S -- \emph{Synthetic DAGs of CyberShake, Epigenomics, LIGO, Montage, SIPHT} \\
			\hline
		\cite{mao:sc:2011} & C & HT & OD & BI-60~min, AL
			& DAG & \{M, HO, T\}/S  & Non-deterministic runtime
			& TD/S & MIN(C) & DAG, Batch
			& Divide and Conquer & Dynamic & S -- \emph{Pipeline, Parallel, Hybrid DAGs} \\
			\hline
		\cite{oprescu:cloudcom:2010} & C & HT & OD & BI-60~min, AL 
			& Task & \{M, HO, B\}/U & Resubmit, Non-deterministic runtime
			& CB/S & MIN(T) & Batch, Batch
			& Dynamic Programming & Dynamic & S -- \emph{BoT w/ Normal Distribution} \\
			\hline

		\cite{poola:iccs:2014} & C & OD:HT, SP:HO & OD, SP  & BI-60~min, AL, PV
			& DAG & S & Checkpoint, Data transfer time
			& TD/S & MIN(C) & DAG, DAG
			& Heuristic & Dynamic & S -- \emph{Synthetic DAG of LIGO} \\
			\hline
		\cite{poola:acm:2016} & C & HT & OD, SP  & BI-60~min, AL, PV, Failures
			& DAG & S & Resubmit, Replicate, Data transfer time
			& TD/S & MIN(C) & DAG, DAG
			& Heuristic & Dynamic & S -- \emph{Synthetic DAG of LIGO}\\
			\hline
\multicolumn{14}{l}{$^\dagger$~Tasks of a BoT are submitted in a batch and scheduled as an ordered collection}\\
\multicolumn{14}{l}{$^{\ddagger}$~Multiple BoTs are submitted transactionally and each BoT is scheduled as a single unit as soon as it arrives}\\

\end{tabular}
\end{sidewaystable*}

\begin{sidewaystable*}
	\centering
	\scriptsize
	
	{\small Table 1. Classification of Scheduling Algorithms \emph{(continued)}}\\

\begin{tabular}{p{0.7cm}||p{0.7cm}||p{0.95cm}|p{0.6cm}|p{1.5cm}||p{0.65cm}|p{1.2cm}|p{1.8cm}||p{1cm}|p{1.4cm}|p{0.8cm}||p{1.6cm}|p{1.0cm}|p{2.2cm}}

		\hline
		\textbf{Paper} 
		& \multicolumn{4}{c||}{\textbf{System Design}~\S~\ref{sec:tax:sys}}  
		& \multicolumn{3}{c||}{\textbf{Application Model}~\S~\ref{sec:tax:app}} 
		& \multicolumn{3}{c||}{\textbf{Quality of Service}~\S~\ref{sec:tax:qos}} 
		& \multicolumn{3}{c}{\textbf{Algorithm}~\S~\ref{sec:tax:algo}} 
		\\ \hline
		& Layer$^1$ & Size$^2$ & Price$^3$ & Char.$^4$
		& Struct. & Mode$^5$ & Char.
		& Constr.$^6$ & Goal$^7$ & Gran.$^8$ 
		&  Technique & Schedule & Evaluation$^9$ \\ \hline
		\hline
		\cite{rodriguez:tcc:2014} & C & HT & OD & BI-60~min, AL, PV
			& DAG & S & Non-deterministic runtime, Data transfer time
			& TD/S & MIN(C) & DAG, DAG
			& PSO & Static & S -- \emph{DAGs of CyberShake, LIGO, Montage, SIPHT} \\
			\hline
		\cite{subramanya2015spoton} & C & HT & OD, SP & BI-60~min, AL
			& Task & S & Resubmit, Replicate, Checkpoint, Migrate
			& -- & MIN(C) and MAX(O) & --, Task
			& Heuristic & Dynamic & S -- \emph{Random tasks, Google cluster trace on EC2} \\
			\hline
			
		\cite{bots-on-ondemand-vm-2} & C & HT & OD & BI-60~min, different PPR, AL 
		& Task & \{M, HO, B\}/U & --
		& TD/H or CB/H & MIN(C) or MIN(T) & Batch, Batch 
		& Heuristic & Static & S -- \emph{Random DAG} \\
		\hline
		\cite{voorsluys:aina:2012} & C & HT & SP & BI-60~min
		& Task & \{M, HO, T\}/O & Replicate, Checkpoint, Migrate, Data transfer time
		& TD/S & MAX(O) and MIN(C) & Task, Batch 
		& Heuristic & Dynamic & S -- \emph{Synthetic tasks based on job stream from LHC Grid} \\
		\hline
		\cite{wu:2013} & C & HT & OD & BI-per unit time, PV w/ bounded randomness, PPR
		& DAG, Task & Service \{M, HO, T\}/S; Task \{M, HT, T\}/U & --
		& TD/H and TB/H & MIN(C) or MIN(T) or (MIN(C) and MIN(T)) & DAG, Batch 
		& Service-- Heuristic; Task-- GA, PSO, ACO & Static+ dynamic & S -- \emph{Random DAGs} \\
		\hline
		\cite{xu:ispa:2009} & C & HT & OD & Limited avail\-abi\-li\-ty
			& DAG & \{M, HO, T\}/O & --
			& TD/S and CB/S & MIN(T) and MIN(C) and MAX(O) & DAG, Batch 
			& Divide and Conquer & Dynamic & S -- \emph{Random DAGs}\\
			\hline
		\cite{zeng:aina:2012} & C & HT & OD & BI-60~min, AL
		& DAG & S & Data transfer time
		& CB/H & MIN(T) &  DAG, DAG
		& Greedy & Static & S -- \emph{Random DAGs}\\
			\hline
		\cite{zhou:tcc:2014} & C & HT & OD & BI-60~min
		& DAG & \{M, HO, T\}/O & Checkpoint
		& TD/S or CB/S & MIN(C) or MIN(T) & DAG, Batch
		& Greedy & Static+ dynamic & S -- \emph{Synthetic DAGs of LIGO, Montage on EC2, Rackspace}\\
			\hline

\end{tabular}
\end{sidewaystable*}

\begin{sidewaystable*}
	\centering
	\scriptsize

	{\small Table 1. Classification of Scheduling Algorithms \emph{(continued)}}\\

\begin{tabular}{p{0.7cm}||p{0.7cm}||p{0.95cm}|p{0.6cm}|p{1.9cm}||p{0.65cm}|p{1.1cm}|p{1.8cm}||p{1cm}|p{1cm}|p{0.8cm}||p{1.6cm}|p{1.0cm}|p{2.2cm}}

		\hline
		\textbf{Paper} 
		& \multicolumn{4}{c||}{\textbf{System Design}~\S~\ref{sec:tax:sys}}  
		& \multicolumn{3}{c||}{\textbf{Application Model}~\S~\ref{sec:tax:app}} 
		& \multicolumn{3}{c||}{\textbf{Quality of Service}~\S~\ref{sec:tax:qos}} 
		& \multicolumn{3}{c}{\textbf{Algorithm}~\S~\ref{sec:tax:algo}} 
		\\ \hline
		& Layer$^1$ & Size$^2$ & Price$^3$ & Char.$^4$
		& Struct. & Mode$^5$ & Char.
		& Constr.$^6$ & Goal$^7$ & Gran.$^8$ 
		&  Technique & Schedule & Evaluation$^9$ \\ \hline

		\hline
		\cite{skarlatprovisioning} & F, C & HT & F:FR, C:OD & BI-60~min, F - Limited availability
		& Task & \{M, HO, B\}/U & Data transfer time
		& -- & MAX(O) and MIN(T) & --, Batch
		& Heuristic & Static & S -- Same length tasks \\ \hline
		\cite{skarlat2017towards} & F, C & HT & F:FR, C:OD & BI-60~min, AL, F - Limited availability
		& DAG & \{M, HO, T\}/U & Data transfer time
		& TD/H & MAX(O) & DAG, Batch
		& ILP & Static & S -- \emph{Sense-process-actuate applications} \\ \hline
		\cite{deng2016optimal} & F, C & F:HT, C:HO & -- & F,C - Limited availability & Task & \{M, HO, T\}/U & Data transfer time & TD/H & MIN(E) & Batch, Batch & Approximation & Static & S -- \emph{Varying number of tasks}\\ \hline
\cite{ghosh-2018} & E, C & HT & -- & E,C - Limited availability & DAG & \{S, HT, B\} & Data transfer, Streaming I/P, Src/Sink pinned & E/H & MIN(T) & DAG & GA, Brute Force & Static & S -- \emph{Real bench\-marks, CEP queries, Random DAGs}\\ \hline
		\cite{ghosh2018adaptive} & E, C & HT & -- & E,C - Limited availability & DAG & \{M, HO, T\}/U & Migrate, Data transfer time & E/H & MIN(T) & Batch, Batch & Heuristic, GA & Dynamic & S -- \emph{Random DAGs} \\ \hline
		\cite{chun2011cloneCloud} & E:M, C & -- & -- & E - Limited availability & DAG & S & Checkpoint, Migrate, Data transfer time & -- & MIN(E) or MIN(T) & --, DAG & ILP & Dynamic & E -- \emph{Virus scanning, image search and behavior profiling applications} \\ \hline
		\cite{shi2012serendipity} & E:M & HT & -- & AL (Deferred), Limited availability & DAG & \{M, HO, T\}/O & Resubmit, Migrate, Data transfer time & TD/H & MIN(E) or MIN(T) & DAG, MIN(E)-Batch; MIN(T)-DAG & Greedy & Dynamic & S, E -- \emph{Face detection and speech to text applications} \\ \hline
		\cite{hong:mcc:2013} & E:M, F, C & HT & -- & E - Limited availability & DAG & S & Migrate & S/H & MIN(T) & DAG, DAG & Heuristic & Dynamic & S -- \emph{Vehicle-to-vehicle video streaming and mobile CEP applications} \\ \hline
		\cite{brogi2017qos} & F, C & HT & -- & F - Limited availability & DAG & S & -- & S/H & -- & DAG, -- & Heuristic, Backtracking, Greedy & Static & S -- \emph{Fire alarm application} \\ \hline
\cite{URGAONKAR2015205} & E:M, C & HT & NP & E-VM migrate, Time slots, Locality & Task & \{S, HT, T\}/O & Locality, Tasks on specific Edges & TD & MIN(O): NW use & Task & Markov Decision Process/ILP & Static & S -- 
\emph{Real mobility traces, cellular NW} \\ \hline
  \cite{bittencourt2017mobility} & F, C & HO & -- & F - Limited availability, locality & DAG & \{S, HT, T\} & DAG locality, Data transfer time & -- & MIN(T), MIN(O): NW use & DAG & Locality-aware, FIFO, Heuristic & Dynamic & S -- \emph{iFogSim, gaming, video processing} \\ \hline

		\end{tabular}
	\end{sidewaystable*}

\section{Classification of Scheduling Algorithms using Taxonomy}
\label{sec:table}

Table~\ref{tbl:review:1} classifies $36$ key publications on application scheduling, based on the taxonomy we have proposed. $25$ of these papers propose scheduling exclusively on cloud resources, while $11$ use a mix of edge, fog and/or cloud. 
The skew toward cloud-based scheduling reflects the decade-long period of maturity of this resource, relative to edge and fog resources that are more recent.
The footnotes of the table indicate the short-hand of the taxonomy terms used within the columns for brevity. 

In summary, on the \emph{system model} dimension, $23$ of the $25$ papers consider cloud infrastructure with heterogeneous VM sizes, $17$ use on-demand VMs with fixed prices, $2$ use pre-emptive VMs, and $5$ use a mix of both. For the mix of edge, fog and cloud, almost all the papers have considered heterogeneous resources and only $1$ paper considers homogeneous cloud resources.

A majority of the cloud scheduling publications -- $19$ to be exact -- use hourly billing that was common among cloud providers, while $2$ others use more current fine-grained billing intervals. Further, $12$ of the papers cataloged consider VM acquisition lag in their schedule planning. Three of these papers also constrain the availability of resources from the provider, which is more representative of Grid and HPC systems than public clouds which provide seemingly unlimited resources on-demand. 
For edge, fog and cloud resources, most of the papers do not consider resource pricing at all. Only $2$ papers consider hourly billing and that too only for the cloud resources. Also $1$ paper has considered deployment lag and $1$ has considered deferred acquisition. All these papers constrain the availability of edge and fog resources, either on account of mobility or limited resources capacity, while $2$ limit the number of available cloud resources as well. $3$ papers consider spatial mobility of edge and fog resources, while two others are have spatial locality between the client and the edge or fog.
 
The reviewed papers have diverse \emph{application characteristics} as well.
Out of all the papers, $25$ use a DAG structure while $12$ use a task structure; $7$ publications allow batch submission while $14$ have a transactional mode of arrival; and $15$ consider applications that are scheduled as a collection rather than individually. As many as $13$ of the papers use time deadline as a constraint and cost as the optimization goal when specifying their \emph{QoS} requirements. However, $8$ articles use cost budget as a constraint as well. Most of the papers that include edge and fog resources aim at minimizing the power consumption, overall latency or network usage.

Most of the research papers use a heuristic or greedy \emph{algorithm for scheduling}, of which $16$ schedule the applications statically, $19$ schedule them dynamically while $2$ use a mix of both approaches. Simulations based on synthetic DAGs or tasks tend to be the predominant means of evaluating these scheduling algorithms, though $5$ of the papers reviewed use real or realistic applications for validation.

This diversity in the surveyed literature indicates the rich classes of problem definitions and corresponding research outcomes in this area of application scheduling on edge, fog and cloud. It also justifies the need for a detailed taxonomy such as ours to categorize and analyze this body of work -- both what has been done and future work that can benefit from these learnings, particularly for a mix of these resource abstractions.

\section{Related Work}
\label{sec:related}
Several existing literature offer generic reviews of edge, fog or clouds, either independently or in comparison. They also offer complementary aspects of application models on each resource independently, or scheduling for other levels of the cloud abstractions. But none offer a holistic survey on the approaches to application scheduling on these distributed infrastructure, with structure and rigor as we have presented here.

\subsection{Resource Characterization}
There are several surveys on the characteristics of edge, fog and cloud resources. These describe the capabilities of the individual resource platforms, and their features that benefit both end users and service providers. This is necessary to examine application scheduling, but needs to be substantiated with applications models and scheduling techniques that leverage these features.

A recent manifesto on cloud computing offers an \emph{overarching review} of contemporary cloud capabilities, and future potential and challenges~\cite{manifesto}. Our cloud resource characteristics are not as detailed, but focus on specific features of relevance to application scheduling. \cite{coutinho:2015} offers a more specialized bibliometric survey on \emph{elasticity of cloud resources}, along with the journals where the publications appear, country of origin of authors, and year of publication. 
A taxonomy of methods used to leverage as well as QoS metrics is also provided. While some of our resource characteristics, application characteristics and QoS topics overlap with this, we do not exclusively focus on elasticity but rather examine other dimensions of clouds such as pricing and resilience as well, in addition to application scheduling.

Similarly, there have been several papers that conceptualize the idea of fog computing and Cloudlets~\cite{varshney2017icfec,Bonomi2014,satya:pervasive:2009}.

Others examine specific applications that benefit from the fog layer, to complement edge and cloud computing, with smart cities and IoT being key drivers~\cite{dastjerdi:computer:2016,yannuzzi:camad:2014,stojmenovic:fedcsis:2014}. Some also prescribe different types of interactions between the edge, fog and cloud layers~\cite{stojmenovic:2014}. These are similar to our resource taxonomy, but fail to compare common and contrasting features across edge, fog and cloud, and how application models and schedulers leverage them.

There are articles which discuss the use of edge, fog and cloud layers in a hierarchical architecture. ~\cite{7721750} consider mobility to be a characteristic feature of both the edge and fog layers, with challenges to resource discovery, service scheduling, QoS guarantee and security. 
A medical emergency use case is used to illustrate the relative benefits of using a cloud-only application deployment design, with one that uses all three. 
Latency is seen as a QoS goal, but they omit concerns on monetary cost and energy usage. 
Similarly~\cite{8123913} 
highlight the security and privacy issues in including edge resources for storage and computation of IoT applications, besides the cloud. 
A fog layer is not explicitly considered, and resource pricing is mentioned in passing. 

The benefits for different IoT applications such as smart grid, smart city and smart transportation are mentioned. 

Our survey goes beyond exemplars and considers how generic application can be conceptually specified in terms of their structure, mode of submission and QoS. We also provide a categorization of possible scheduling algorithms in literature. Individual and combined resource layers are included in our review, without limiting to a hierarchical model.

\emph{Mobile Edge Computing (MEC)} has received particular attention due to early mobile phones that were resource constrained.  
~\cite{7879258} surveys the existing research on computation offloading in MEC and how it is integrated with the mobile network architectures. 
They illustrate use cases that benefit from MEC and their application characteristics, such as whether the application can be partitioned and offloaded, dependencies between parts, predictability of the input data size. 
These affect how an application can be executed locally on the phone, partially offloaded or fully offloaded to the cloud 
to meet the QoS goals, such as minimization of delay or energy consumption.  
Techniques for responding to the device's mobility, like changing transmission power, VM migration, and communication route selection are suggested. 
~\cite{8030322} 
offers a conceptual model of computing that moves between a mobile edge device and the cloud, with the 

intelligence on application scheduling present between them on the Radio Access Network (RAN).  
This communication-centric view considers fog and edge computing as the same resource layer. 
They review literature on computation offloading in MEC, with low latency processing, storage issues, and energy efficiency being QoS goals. 
Our survey while similar in spirit to these 
takes a more holistic and forward-looking view of the resource characteristics of edge, fog and cloud, including pricing and performance, and generalizes the application structure and their QoS.

\subsection{Scheduling Techniques}

Scheduling for different aspects cloud computing have been examined in the past. \cite{zhan:csur:2015} present a taxonomy of \emph{scheduling algorithms at various layers}: application, virtualization and infrastructure. Algorithms are classified based on the specific objectives that should be met at each layer. Scheduling at the virtualization layer has the goal of mapping VMs on to physical machines, which is of particular interest to service providers. This been addressed by other specialized surveys such as \cite{pietri:2016,mann:csur:2015}.  The goal of scheduling at the infrastructure layer is to place the resources and services at different locations in various data centers. 
\cite{grozev:spe:2014} drill-down into this layer and offer a taxonomy for inter-cloud architectures, with application brokering in these systems. 
At the application layer, the problem is to schedule the user's application on VMs, and three goals are discussed by \cite{zhan:csur:2015}: user QoS, provider efficiency, and negotiation. 

In contrast, our survey focuses on just the application layer but goes in-depth by offering a detailed classification 
of edge, fog and cloud resources based on pricing models and system characteristics essential for scheduling algorithms. Moreover, we also categorize the structure of workflows, their characteristics and mode of submission. We further identify fault tolerance as a user QoS with scheduling techniques designed for application resilience on these resource layers.

Our work is of broader interest to middleware and application developers using these distributed resources, 
who, arguably form a larger population that can put this survey to practical use, compared to commercial resource service providers who are less influenced by such research outcomes.

Similarly, \cite{kessaci:ipdpsw:2014} also discuss \emph{scheduling at the three levels} -- service, task and VM -- for public and private clouds. We limit our focus to \emph{public clouds} that are exceedingly popular and where issues of elasticity and costing offer clear challenges and opportunities to end users, and complement this with emerging edge and fog resource layers. 
Our work also emphasizes the \emph{application layer}, a subset of which is the task level considered by~\cite{kessaci:ipdpsw:2014}, with the objective of meeting QoS requirements and/or budget constraints for the application. We also offer greater detail on various aspects of the unit of scheduling, 
which is lacking in these surveys.

Likewise, \cite{huang:jsw:2013} limits itself to resource allocation in clouds in the presence of \emph{system failures}, with various ACO and GA based dynamic scheduling meta-heuristics to handle faults being explored. 
We go beyond system failure and also examine literature that handle failures due to pricing (out of bid) for pre-emptible and opportunistic resources. 

We also discuss several other scheduling algorithms that address diverse QoS constraints and goals.

\cite{wu:2015} presents several \emph{categories of workflow scheduling algorithms} on clouds, including static and dynamic scheduling, to meet constraints such as budget, deadline and robustness.  

Similarly, \cite{liu:ccgrid:2014} offers a brief summary of workflows and their objective criteria for scheduling on the cloud, and review literature on scheduling algorithms to achieve the same.  
Others have surveyed \emph{meta-heuristic techniques} for scheduling workflows on the cloud~\cite{tsai:jsyst:2014}. These techniques include hill climbing, simulated annealing, tabu search, GA, PSO and ACO, and these have been used to generate a schedule for workflows on clouds. 
While these surveys consider multiple classes of scheduling algorithm and system characteristics that lie at the intersection of several dimensions we introduce, they do not offer a taxonomy of the dimensions themselves which is essential for a rigorous analysis of this space. Edge and fog resources which have gained prominence off-late are omitted as well.

Some early research investigates platform and application models for edge and fog computing~\cite{varshney2017icfec}. ~\cite{hong:mcc:2013} propose a 3-level strictly hierarchical model where the computation is rooted in the cloud, resources are elastically acquired in the cloud and fog layers, and communication is possible between cloud and fog, or fog and edge. 
But their example applications do not use the cloud, and 
this degenerates to a client-server model between the edges and their fog parent. 
The role of virtualization in enabling cloud computing is discussed in~\cite{bittencourt:pgcic:2015}, and they see a similar role for the fog as well. They conceive of a VM encapsulating all dependencies for an edge application or user to be hosted on a Cloudlet within 1 hop of the edge, with this VM migrating to remain at 1-hop distance from the edge user. 

Such articles offer potential architectures for interactions between edge, fog and cloud, while out survey more broadly characterizes these resources and examines their impact on applications and how they are scheduled.

\section{Discussion}
\label{sec:discussion}
This detailed review of application scheduling characteristics on edge, fog and cloud resources, along with the feature matrix, highlight several open problems, whose solutions require a mix of research, development and business models. There are also rapidly emerging technologies that can influence these directions. We discuss these along similar categories that we have proposed.

\subsection{Evolution in Resource Abstractions}

Public cloud providers are highly agile and respond rapidly to evolving technologies and market dynamics. 

One challenge is being addressed is \emph{light-weight application sandboxing}, in contrast to hypervisor based VMs. 
Cloud providers like Amazon EC2 and Microsoft Azure are offering Docker containers, which have minimal resource overheads and offer rapid instantiation 
compared to virtualization, and are useful when OS heterogeneity is not required. They even support basic migration between hosts. 
These containers however use kernel-based controls to enforce security and resource allocation, which are less effective than hardware virtualization. This can open up opportunities for trade-off between the ability to respond rapidly to application dynamism, multi-tenant container security, and handling performance variability.

Recent work on minimalist OS like \emph{VMWare's PhotonOS}, light-weight virtualization like \emph{AWS Firecracker}, and even web-standards like \emph{W3C's WebAssembly} offer alternatives to containers and hypervisors. Some of these are motivated by the popularity of a serverless \emph{Function-as-a-Service (FaaS)} model, based on stateless micro-services that encapsulate user-defined functions that can be composed and executed~\cite{akkus2018sand}. This is similar to task scheduling in a transactional model. Besides cloud data centers, FaaS is also being pushed to the fog and edge layers using SDK's like \emph{Amazon Greengrass} and \emph{Azure Edge IoT} offered by the cloud providers. 
These cloud fabrics extend to edge devices, and allow for more centralized management of distributed edge and fog resources on the wide area network~\cite{azureiot,awsgreengrass,manifesto}. These are currently limited to running applications using a FaaS model, with their scheduling managed exclusively by the provider. But the ability to expose IaaS resources on the edge and fog can help further leverage scheduling designed for the cloud to be extended to the edge and fog, and ease the design of practical edge, fog and cloud applications.  

Besides the push of existing cloud providers, there are also alternative \emph{business and technology models for edge and fog computing} that can be sustainable~\cite{vaquero2014finding}, both for infrastructure deployments and platform support~\cite{varshney2017icfec}. 

They are more obvious in vertically integrated ``private'' scenarios rather than horizontal, reusable ``public'' ones, much in the way of cloud data centers evolving from private use by Amazon, Google and Microsoft to a commercial business model of a public cloud~\cite{yi:mobidata:2015,manifesto}. 
Potential providers of 

on-demand public fog computing 
are operators of cell-phone towers who have captive power, communications and space, and 
Smart Cities deployments with captive compute capacity as part of the city deployments of verticals like 
smart power grid or smart transportation
~\cite{stojmenovic:fedcsis:2014,Bonomi2014,amrutur:testbed:2017}. 

Likewise, the advent of energy-efficient and high-performance \emph{accelerators} like GPUs and Tensor Processing Units (TPUs), and low-power ARM64 servers as cloud and fog resources, introduces further resource diversity and application opportunities that impacts scheduling~\cite{kalyanasundaram:hipc:2017,8587006,manifesto}. Part of this is driven by the rapid adoption of machine learning models and deep neural networks, which analyze multimedia data (e.g., video surveillance from smart cities) and have high computing costs~\cite{8342120}. Novel edge and fog devices such as drones and other autonomous vehicles are starting to become a reality as well~\cite{Sathiaseelan:2016:CMC:2935620.2935625}, and introduce new challenges in the energy and compute-constrained mobile resources with transient communications. Likewise, the adoption of 5G communication technology can translate into wide-spread deployment and accessibility of edge and fog devices, offering a high-bandwidth and pervasive last-mile link~\cite{taleb2018survey}. These technology-shifts will require us to revisit many of the assumptions on the system models used for scheduling.

At the same time, there is also a lack of \emph{standardized infrastructure and platform interfaces} for edge and fog computing, with much of the advances in fabric management, programming models, power and network management, fault tolerance and pricing models being limited to research prototypes
~\cite{varshney2017icfec}. 
However, initiatives such as OpenFog Consortium (which recently merged with the Industrial Internet Consortium)~\cite{ofra} and EdgeX Foundry~\cite{edgexfoundry} championed by various industries are starting to offer reference architectures and software stacks to address this gap. These will serve as the vehicle to incorporate and enact the scheduling models that are developed.

\subsection{Application Models and QoS}
Application models tend to evolve more slowly than hardware and communications technologies, which are driven by the industry.
Research has tended to focus more on batch execution of workflows as the unit of scheduling. However, the growth in streaming data and online decision-making applications means that \emph{transactional workloads} and \emph{event-driven} models need to be better examined. 

In fact, processing streaming data and having a control-loop between sensors and actuators, with analytics scheduled in-between, is a common pattern in IoT applications~\cite{simmhan:iotn:2017}. A few of the literature we have tabulated consider such an event-driven or transactional model~\cite{URGAONKAR2015205,ghosh-2018}.
Similarly, \emph{BoTs} are a common abstraction that are inadequately examined for scheduling, even as they make good candidates for off-loading to the cloud pr fog partitions for execution~\cite{ibarra:jacm:1977,varshney:tpds:2018}. The popularity of FaaS also pushes data to the compute, rather than the typical model of moving compute to the data, while easing weak-scaling of stateless micro-services~\cite{akkus2018sand}.

The increasing importance of \emph{machine learning applications} means that scheduling that is sensitive to the goals of training and inferencing will be beneficial~\cite{Zhang:2017:SQS:3127479.3127490,ananthanarayanan2017real}. In particular, the QoS goals and constraints may need to include the quality of the training and inferencing accuracy as first-class metrics, besides the time taken. Further, with increasing \emph{personally identifiable data} being collected and processed from the edge, privacy and trust start playing a key role. This may impose limitations on where to run what applications, and if additional operations such as anonymization or masking is required before placing specific tasks on specific resources~\cite{Wang:2017:SPI:3083187.3083192}. This can require the application to be re-composed on the fly before being scheduled.

Scheduling across \emph{different resource layers} also introduces independent constraints and priorities on each layer that need to be met. E.g., 

energy is a key concern for scheduling on edge and fog, while pricing and latency are user concerns when using the cloud.  
Additional research is required into capturing these resource-specific factors into the optimization goals or constraints.

\subsection{Scheduling Approach}
While there has been a lot of conceptual work on making use of edge, fog and cloud resources, there is a lack of literature on novel scheduling approaches that consider all the $3$ layers together while leveraging their relative merits. 
Further, the interactions between the layers currently tends to be limited to a hierarchy of cloud--fog--edge, or just a flat logical structure composed of heterogeneous resources across them.

This is a need to examine distributed scheduling strategies rather than centralized ones to ensure scaling to thousands of resources on the local and wide area network, and also resilience to avoid a single point of failure. Techniques from P2P can play a role here~\cite{lo2004cluster}.

Given the challenges in access to large scale edge and fog deployments, much of the validation of scheduling approaches for these resources are based on simulations, using frameworks like iFogSim~\cite{fogsim}. However, there has been work on \emph{virtual environments} like VIoLET~\cite{violet} that use containers running on VMs or clusters to replicate the behavior of edge, fog and cloud resources using container resource allocations. It can also control the network topology between the resources, and the resource dynamism and failures. These have the benefit of allowing real applications to be scheduled, executed and the schedule evaluated for realistic edge, fog and cloud resources, and offer a balance between empirical and simulation-based approaches. More such efforts are required to model communication diversity, device reliability, etc.

\section{Conclusion}
\label{sec:conclude}

In this survey, we have offered a comprehensive taxonomy for defining and designing application scheduling algorithms on edge, fog and cloud infrastructure, based on a detailed literature review. These span the system model for the three layers, application model with an emphasis on DAG and task-based applications, and the QoS goals and constraints to be met, which collectively help define scheduling as an optimization problem. We also categorize existing approaches to solving this optimization problem and evaluating it, based on prior reviews. This taxonomy has been used to tabulate the characteristics of $36$ research papers on scheduling on 
edge, fog and/or cloud resources. 
The taxonomy presents \emph{designers} of scheduling algorithms for edge, fog and cloud applications with a clear set of system and application features they should consider for their target infrastructure and workload. The table provides \emph{architects} of application runtimes with the relevant classes of scheduling algorithms that they can leverage to meet the needs of their end-users. 
Learning from this body of work is essential as applications are increasingly designed for edge and fog environments, rather than just the cloud, and scheduling challenges are set to emerge within newer application platforms that can operate on these resources. 
To this end, we have also explored various emerging technology trends that will require us to re-examine current system and application models, and develop novel scheduling techniques for the next generating of computing.

\bibliographystyle{plain}
\bibliography{references}

\appendix
\section*{Appendix}
\begin{sidewaysfigure*}

	\centering
	\includegraphics[width=1\textwidth]{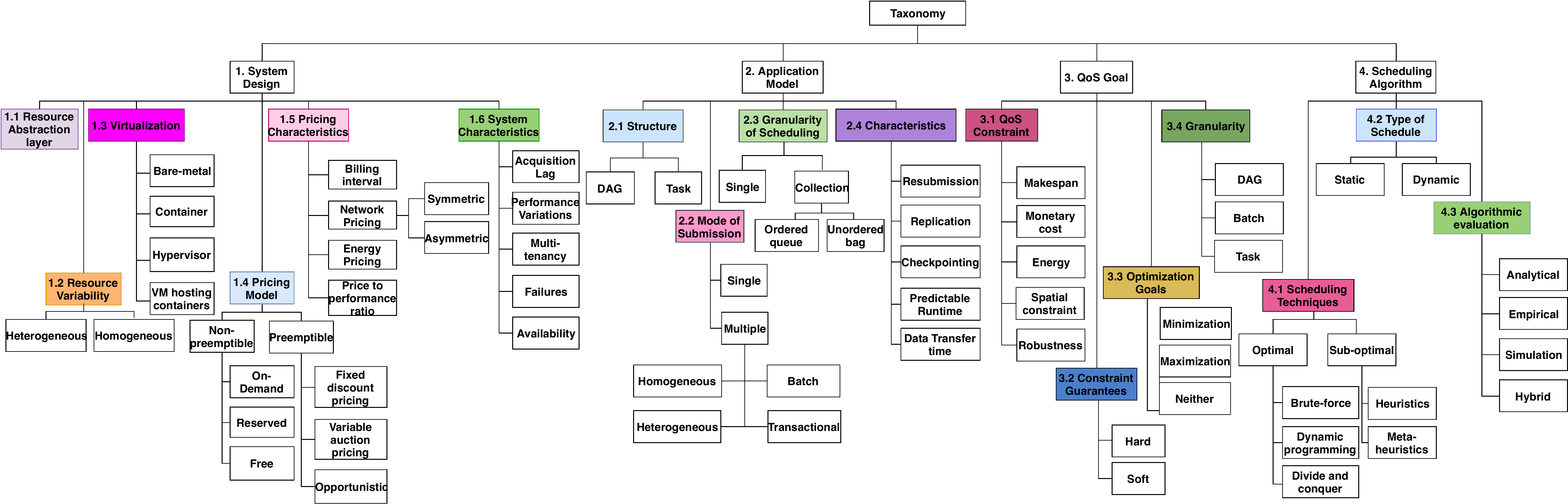}
	\caption{Complete Taxonomy as a Single Tree}
	\label{fig:tax:full}

\end{sidewaysfigure*}

\end{document}